\documentstyle[aps,epsfig]{revtex}
\begin{document} 
\draft
\title{Comments on the renormalizability of the Broken Symmetry Phase in
Noncommutative Scalar Field Theory }
\author{S. Sarkar and B. Sathiapalan} 
\address {The Institute of Mathematical Sciences, C. I. T. Campus, Chennai-600 113, 
India.}                                                          
\date{\today}
\maketitle 
\begin{abstract}
We study the noncommutative $\phi^4$ theory with spontaneously broken global O(2) symmetry in 4
dimensions. We demonstrate the renormalizability at one loop. This does
not require any choice of ordering of the fields in the interaction terms.
It involves regulating the ultraviolet and infrared divergences in a manner consistent with the Ward identities.
\end{abstract}

\section{introduction}
Noncommutative spacetimes have been studied intensively over the last few
years \cite{Connes}\cite{bal}\cite{witten}. The fact that they occur as
low energy limit of String
Theory when a constant $B_{\mu\nu}$ background is turned on makes the
subject even more interesting \cite{witten}. A perturbative study of $\phi^4$ field
theory revealed an intriguing transmutation of some UV divergences into IR
divergences \cite{min}. This is very reminiscent of what happens in string
perturbation theory where the world sheet duality or modular invariance
maps the UV region into the IR region. This is the origin of the well
known fact that there are no UV divergences in String Theory.

In the light of the above remarks it is natural to wonder about
perturbative renormalizability of the noncommutative $\phi^4$ theory. It
was argued in \cite{min} that the UV divergences that continue to
arise in the theory are from Planar graphs \cite{filk} and are exactly of
the same
form as in the commutative theory and hence can be renormalized away. The UV
divergences from the non-planar graphs are, on the face of it, more
severe. However they turn into IR divergences in an intriguing way and are
UV finite. The non-planar graphs therefore do not threaten the
renormalizability of the theory. Various other aspects of Noncommutative
field theories have been studied in [6-34].

Issues concerning the renormalizability of theories with global
O(N) symmetry,
in the large N limit have been addressed in \cite{gubser},\cite{emil}. 
One can also study these issues in theories with global O(2) symmetry, in
their broken/unbroken phases. 
In commutative theories if the theory is renormalizable in the unbroken
phase it is trivially
renormalizable in the broken phase. The Ward Identities guarantee this as
also the masslessness of the Goldstone Bosons, to all orders in
perturbation theory. However it was found in \cite{kirk} that in the
noncommutative O(2) theory there seems to
be a violation of these identities at the one loop level. The crucial
point being that the counterterm that renormalized the $\sigma$-tadpole
divergence could not remove the divergences in the $\pi$-$\pi$ two point
function. It was also found \cite{frank},\cite{kirk} that for a particular
choice of
ordering of fields in the $\phi^4$ couplings, the theory is
renormalizable.

Given that both the path integral measure and the action of the noncommutative
theory have global O(2) symmetry one would expect this symmetry to hold at
one loop (in fact to all orders). We investigate this in this paper. We
show that the apparent violation of the Ward identities can be traced to
the dropping of surface terms in the effective action. These are UV
divergent surface terms. If one modifies the prescription to keep these
terms, the broken theory is renormalizable. The crucial point is to keep
an infrared regulator while the continuum limit ($\Lambda \rightarrow
\infty$) is taken for all the processes in the theory. 
This prescription allows one to maintain O(2) symmetry in the UV
divergent terms. However the infrared
divergences make it difficult to use the Ward Identities for the two point 
function in the limit $p\rightarrow 0$ (which is required to establish the
masslessness of the pion).

This paper is organised as follows. In section II we describe the basic problem 
in the context of an O(2) model and point out the reason for it and suggest
a resolution. In section III as a preliminary step we calculate one loop graphs in the 
symmetric phase and discuss the Ward Identity. In section IV as a next step we expand 
one of the fields about a background value, but only in external legs i.e. masses
are kept at the symmetric (tachyonic) value. We show that one can preserve the Ward 
Identity if we set the momentum of the background to zero only at the end of the 
loop calculations. In section V we consider the broken phase of the theory where the masses in the loop are
the physical (tree-level) masses and describe a prescription that allows the Ward Identities
to be maintained (at one loop). We give our conclusions in section VI.

\section{ Ward Identities and outline of the prescription}

In this section we derive the relevant Ward Identities for the symmetric
as well as for the broken phases and show  in the following
sections that the Ward Identities hold for one loop calculations.

Consider the global O(2) invariant action $S[\phi]$,

\begin{eqnarray}
Z[J]&=&\int {\cal D}\phi_1 {\cal D}\phi_2 e^{i[S[\phi] + \int d^4x\phi_i
J_i]}\\ \nonumber
[i &=& 1,2]
\end{eqnarray}

The action is invariant under the following transformation,

\begin{eqnarray}
\phi^i \rightarrow \phi^i + \epsilon^{ij} \phi^j
\end{eqnarray} 

where $\epsilon^{ij}$ is an infinitesimal parameter antisymmetric in the
$i$, $j$ indices. So is the measure defined by,

\begin{equation}
\int {\cal D}\phi e^{[-\frac{1}{2}\int d^4x\phi * \phi]}= \int {\cal D}\phi e^{[-\frac{1}{2}\int
d^4x\phi^2]}=1
\end{equation}

 Where, the $*$-product is defined by,

\begin{equation}
(\phi_1*\phi_2)(x)=e^{\frac{i}
{2}\theta^{\mu\nu}\partial_{\mu}^y
\partial_{\nu}^z}\phi_1(y)\phi_2(z)\mid_{y=z=x}
\end{equation}

$\theta^{\mu\nu}$ is real and antisymmetric.
We have used the fact \cite{min} that in quadratic terms the $*$-product can be replaced by an
ordinary product.
Since action and the measure are invariant under this transformation, the
generating functional satisfies,

\begin{eqnarray}
Z[J]=Z[J+\epsilon J]
\end{eqnarray}
 
or,

\begin{eqnarray}
\epsilon^{ij} J^i \frac{\delta Z}{\delta J^j}=0
\end{eqnarray}

Using the definition of quantum effective action,

\begin{eqnarray}
\Gamma[\phi] = -\int d^4x \phi^i(x) J^i(x) + W[J]
\\ \nonumber W[J]=-ilnZ[J]
\end{eqnarray}

We have the following set of Ward Identities,

\begin{eqnarray}
\phi_1\frac{\delta \Gamma}{\delta \phi_2}&=&\phi_2\frac{\delta
\Gamma}{\delta \phi_1}\\
\frac{\delta^2 \Gamma}{\delta \phi_{1}^2} \mid_{\phi_1=\phi_2=0}&=&\frac{\delta^2 \Gamma}{\delta
\phi_{2}^2}\mid_{\phi_1=\phi_2=0}\\
\frac{\delta^4 \Gamma}{\delta \phi_{1}^4} \mid_{\phi_1=\phi_2=0} &=& 3\frac{\delta^4 \Gamma}{\delta 
\phi_{1}^2 \delta\phi_{2}^2 }  \mid_{\phi_1=\phi_2=0} 
\end{eqnarray}

where, $\frac{\delta^n\Gamma}{\delta \phi^n} \mid_{\phi=0}= \Gamma^{n\phi}$ generates
all
one particle irreducible $n$ point functions.
Now if the symmetry is broken so that $\phi_1=\sigma + v$ and
$\phi_2=\pi$, we get 

\begin{eqnarray}
v\frac{\delta^2\Gamma}{\delta\pi^2} \mid_{\sigma=\pi=0}=\frac{\delta\Gamma}{\delta\sigma}  \mid_{\sigma=\pi=0}
\end{eqnarray}

Equation (10) says that if we set $v$ to a constant equal to the value of
$\phi_1$ at the minimum of the potential then the $\pi$-mass is zero. This
is Goldstone's Theorem. There are Infrared divergences in the
theory of the form $\frac{1}{pop}(\phi_{1}^2+\phi_{2}^2)$ in $V(\phi)$, where $pop=
-\frac{p^{\mu}\theta_{\mu\nu}^2p^\nu}{4}$. Naively this implies that the
symmetric phase vacuum with $\phi=0$ is always a
lower energy solution and therefore at one loop one has symmetry
restoration. We are not sure if this is the right
interpretation. A resolution of this requires a proper treatment of the IR
divergences. We do not have any suggestions for this in this paper.

We consider the noncommutative O(2) model with the
only interaction
term $-\frac{\lambda}{4}(\delta^{ij}\phi^i*\phi^j)^2$. We refer the reader to
\cite{min} for an introduction to Noncommutative field theories. 
As pointed out , if the symmetries of the classical theory are
respected by the quantum theory, all UV divergent terms must be 
O(N) invariant if the UV regulator is chosen
so that it respects this symmetry. We show that to one loop the symmetry is preserved
i.e. the counterterms required to absorb the ultra violet divergences are
O(2) symmetric. 

The terms containing $\exp^{ik\wedge p}$ in the integrals, with $k$ as the loop
momentum
being integrated over, are regulated. We shall call such terms as
Non-Planar(NP) terms. Such terms have $\Lambda$ dependence of the form
$\frac{1}{\frac{1}{\Lambda^2} +pop}$ ,
where $\Lambda$ is the UV cut off. As suggested in \cite{min} we shall treat these terms as potentially
IR
divergent terms when $p\rightarrow 0$.

The Ward Identity, equation (11) shows that the $\sigma$ and the
$\pi$-$\pi$
amplitudes are identical in the broken theory. The Planar and the Non-Planar 
terms must be equal separately for both the amplitudes.  
However it was shown \cite{kirk} that the Planar and the Non-Planar terms
do not match
in the broken theory thus raising the question of renormalizability of the 
theory. In fact naively the $\sigma$ tadpole amplitude does not have any Non-Planar 
term at all. 

We now take another look at the problem.
Let us  expand  the  Non-Planar terms, which contain the factor $\Lambda_{eff}$

\begin{eqnarray}
\nonumber
\Lambda_{eff}^2=\frac{1}{\frac{1}{\Lambda^2} + pop} =\Lambda^2 - \Lambda^4pop + .... \\
ln\Lambda_{eff}^2=ln[\frac{1}{\frac{1}{\Lambda^2} + pop}] = ln\Lambda^2 - \Lambda^2pop + ....
\end{eqnarray}


Note that an infinite number of increasingly UV divergent terms are summed
to give an UV finite term. What is more, when multiplying a field, as in
one of the terms of

\begin{eqnarray}
\Gamma^{\sigma} \sim v_0\int d^4 x \Lambda_{eff}(x)\sigma (x),  
\end{eqnarray}

these are all surface terms. Where $p$ in $\Lambda_{eff}$ is replaced by
$\partial$. The violation of the Ward Identity, (11) is thus due to the
dropping of these surface terms. The $\sigma$ amplitude thus can be
modified and made equal to the $\pi$-$\pi$ amplitude if we retain these
terms.


We now outline a simple prescription that takes care of all these surface
terms. We break the O(2) symmetry by shifting one 
of the fields by an amount $v$ that is not a constant. Thus the Ward
Identity (11) becomes,
\begin{eqnarray}
\nonumber
\int \frac{d^4 p}{(2\pi)^4} v(-p)
\frac{\delta^2\Gamma}{\delta\pi(p_1)\delta\pi(p)}\mid_{\sigma=\pi=0}
=\frac{\delta\Gamma}{\delta\sigma(p_1)}\mid_{\sigma=\pi=0}
\end{eqnarray}
 
We shall treat $v$
as a background classical field and restrict it to a
constant only after all loop computations. With this we would be
taking all Infrared limits of the theory at the same time. This is a form
of infrared regularization. Thus we keep an infrared regulator while the 
continuum limit $\Lambda \rightarrow \infty$ is taken. 
It ensures that the Ward Identities
corresponding to the O(2) symmetry are satisfied. 

In section IV we show how the
renormalizability of this  broken theory
follows from the O(2) symmetry almost trivially. However this theory is 
tachyonic like  the symmetric theory  since the
quadratic terms in $v$ being $x$-dependent are treated as mass insertions 
 and so do not modify the masses of
the resulting fields in the broken phase. In 
section V we show how  we can do computations with the fields having  modified
masses by invoking some additional rules regarding mass insertions in the
graphs that would preserve the division into Planar and Non-Planar
diagrams so as to keep the renormalizability of theory intact.


\section{Symmetric Phase Calculation}

\noindent 
We now compute the one loop amplitudes of the symmetric theory and show that the quantum effective action
is O(2) symmetric to one loop.

The noncommutative O(2) invariant lagrangian is

\begin{eqnarray}
\L_S &=& \frac{1}{2}(\partial_{\mu}\phi^i)^2 + \frac{1}{2}\mu^2(\phi^i)^2 
-\frac{\lambda}{4}\phi^i*\phi^i*\phi^j*\phi^j+{\cal L}_{ct}
\\ 
{\cal L}_{ct}&=&\frac{\delta_{z}}{2}(\partial_{\mu}\phi^i)^2 +
\frac{\delta_{\mu}}{2}(\phi^i)^2 
-\frac{\delta_{\lambda}}{4}\phi^i*\phi^i*\phi^j*\phi^j
\\
\nonumber
i,j &=& 1,2 
\end{eqnarray}

where $\mu^2 > 0$, so the theory is tachyonic in this symmetric
phase. Note that there are several inequivalent orderings of the fields
possible for the quartic term. We have chosen one. Since the quartic term
is O(2) invariant for any ordering one should expect O(2) symmetry to be
preserved at the quantum level also. Furthermore the UV divergences come
from the planar graphs where the ordering is not important. Therefore one
expects the theory to be renormalizable for any choice of ordering.  
We are interested in showing that the quantum theory is
symmetric and that the spontaneously broken theory is renormalizable.
The Feynman rules for the symmetric phase are listed in the appendix. We shall
not be keeping track of factors of $(4\pi)^2$. We shall be
doing all computations in Euclidean space.


\subsection{Two point correlation functions}

\begin{figure}[htbp]
\begin{center}
\epsfig{file=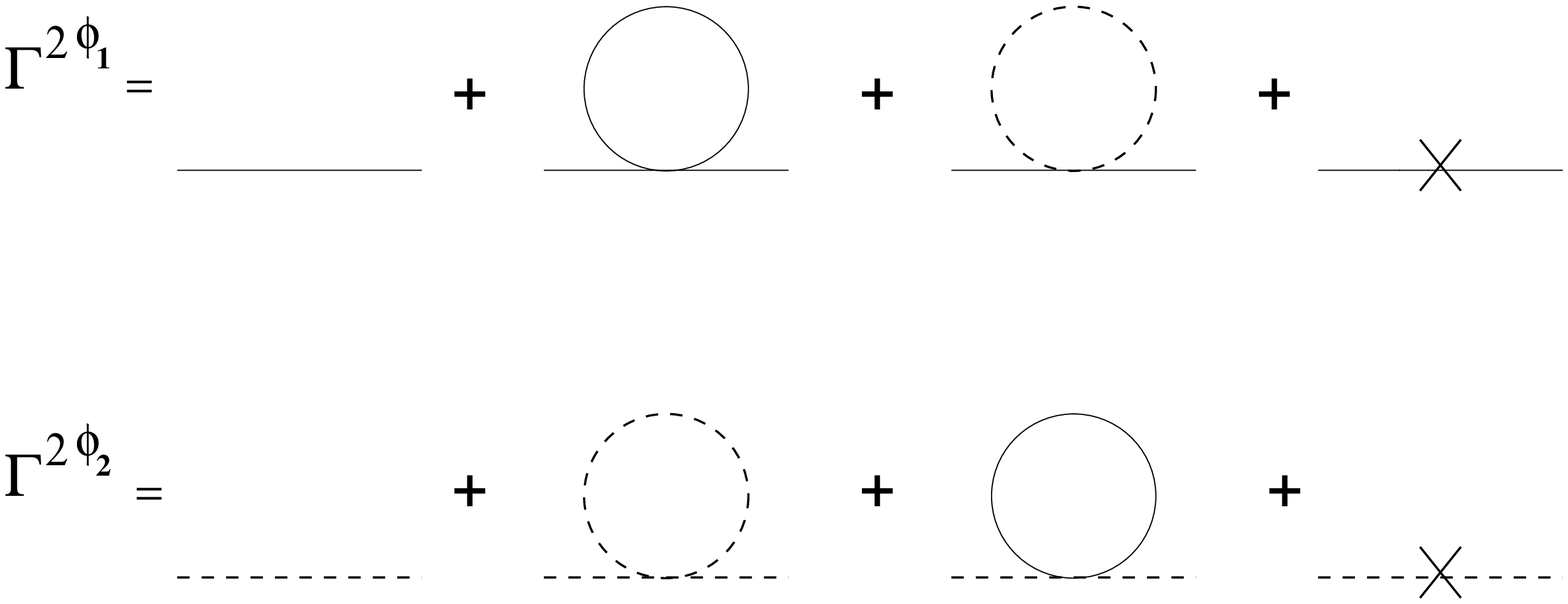, width= 12 cm,angle=0}
\vspace{ .2 in }
\begin{caption}
{}
\end{caption}
\end{center}
\end{figure}

\begin{eqnarray}
\Gamma^{2\phi_1}&=&[-p^2+\mu^2]+\frac{1}{2}(-2\lambda)\int\frac{d^4k}{(2\pi)^4}
\frac{2+cos(p\wedge k)}{k^2-\mu^2}+ 
\frac{1}{2}(-2\lambda)\int\frac{d^4k}{(2\pi)^4}
\frac{1}{k^2-\mu^2}+[-\delta_{z} p^2+\delta_{\mu}]
\\ \nonumber &=& \Gamma^{2\phi_2}
\\ \nonumber &=&
[-p^2+\mu^2]+[-(2+1)\lambda(\Lambda^2+\mu^2 
ln(\frac{\Lambda^2}{-\mu^2}))+NP
+F]+ [-\delta_{z} p^2+\delta_{\mu}] 
\end{eqnarray}

Equation (16) is the expression for the amplitude shown diagramatically in
Fig.1. NP is the Non-Planar ultraviolet finite term regulated by the $cos$ 
present within the
integral and F is the finite term. The contributions from each of the
diagrams have been written
down separately as these terms would be individually required when we show
the renormalizability of the broken theory. It is clear that the quadratic
term in the effective action is O(2) symmetric. In fact the ward
identity, eqation (9) is trivially satisfied.
We now show that the quartic terms are O(2) symmetric as well. The general
four point amplitude for the theory is given in the appendix. Note that diagrams
with more than four external legs are UV finite.


\subsection{Four point correlation functions} 

\begin{figure}[htbp]
\begin{center}
\epsfig{file=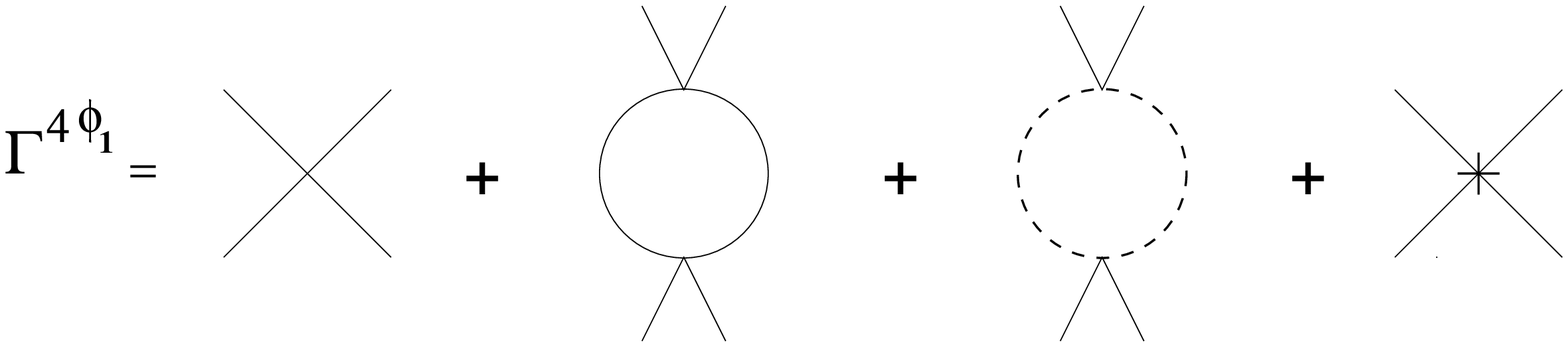, width= 14 cm,angle=0}
\vspace{ .2 in }
\begin{caption}
{  }
\end{caption}
\end{center}
\end{figure}

Taking into account all the three channels (s,t,u) and writing out
explicitly the UV contribution from each process we have,

\begin{eqnarray}
\Gamma^{4\phi_1}&=&-2\lambda f_1(p_1,p_2,p_3,p_4)\\ \nonumber &+&
4\lambda^2(\frac{1}{2})[(2+\frac{1}{2})f_1(p_1,p_2,p_3,p_4)\int\frac{d^4k} 
{(2\pi)^4}\frac{1}{(k^2-\mu^2)[(p-k)^2-\mu^2]}] +NP   
-2\delta_{\lambda}f_1(p_1,p_2,p_3,p_4)\\ \nonumber
&=&\Gamma^{4\phi_2}\\ 
&=&-2\lambda f_1(p_1,p_2,p_3,p_4) + 
2\lambda^2[(2+\frac{1}{2})f_1(p_1,p_2,p_3,p_4)ln(\frac{\Lambda^2}{-\mu^2}) +
NP+ F] \\ \nonumber
&-&2\delta_{\lambda}f_1(p_1,p_2,p_3,p_4)
\end{eqnarray}

where,
\begin{eqnarray}
\nonumber
f_1(p_1,p_2,p_3,p_4)=[cos(p_1\wedge p_2)cos(p_3 \wedge p_4)
&+& cos(p_1\wedge p_3)cos(p_2\wedge p_4)\\ \nonumber 
&+& cos(p_1\wedge p_4)cos(p_2\wedge p_3)]\delta(p_1+p_2+p_3+p_4) 
\end{eqnarray}

\begin{figure}[htbp]
\begin{center}
\epsfig{file=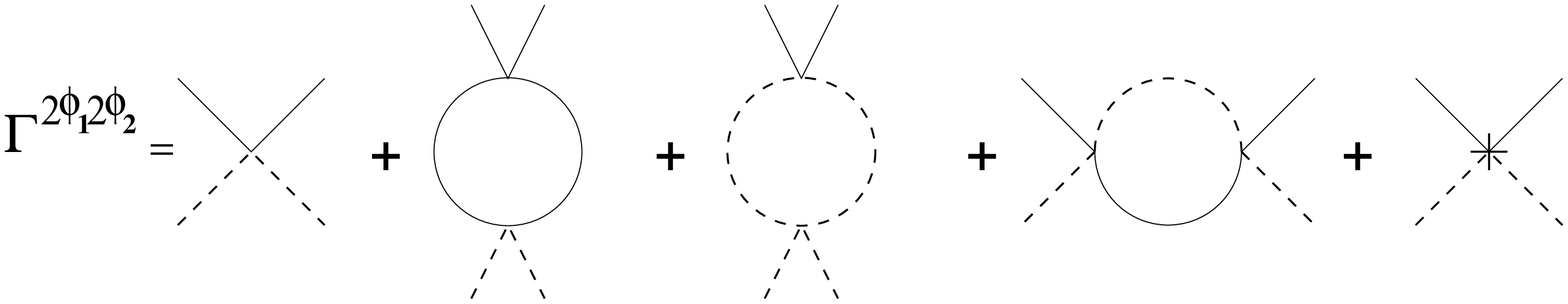, width= 14 cm,angle=0}
\vspace{ .2 in }
\begin{caption}
{   }
\end{caption}
\end{center}
\end{figure}

\begin{eqnarray}
\Gamma^{2\phi_1 2\phi_2}&=&-2\lambda f_2(p_1,p_2,p_3,p_4)\\ \nonumber &+&
4\lambda^2(\frac{1}{2})[(1+1+\frac{1}{2})f_2(p_1,p_2,p_3,p_4)\int\frac{d^4k} 
{(2\pi)^4}\frac{1}{(k^2-\mu^2)[(p-k)^2-\mu^2]}] +NP 
-2\delta_{\lambda}f_2(p_1,p_2,p_3,p_4)\\ \nonumber
&=&-2\lambda f_2(p_1,p_2,p_3,p_4)+ 
[2\lambda^2(1+1+\frac{1}{2})f_2(p_1,p_2,p_3,p_4)ln(\frac{\Lambda^2} 
{-\mu^2})+ NP]\\ \nonumber
&-&2\delta_{\lambda}f_2(p_1,p_2,p_3,p_4)
\end{eqnarray}

where,
\begin{eqnarray}
\nonumber
f_2(p_1,p_2,p_3,p_4)=[cos(p_1\wedge
p_2)cos(p_3\wedge p_4)]\delta(p_1+p_2+p_3+p_4)
\end{eqnarray}

$f_1(p_1,p_2,p_3,p_4), f_2(p_1,p_2,p_3,p_4)$ will stand for these 
expressions everywhere in this paper.

Now, symmetrizing the $\Gamma^{4\phi_1}$, $\Gamma^{2\phi_1 2\phi_2}$
amplitudes with respect to the external momenta $p_1$,$p_2$,$p_3$,$p_4$,
we have,

\begin{equation}
\Gamma^{4\phi_1}=3\Gamma^{2\phi_1 2\phi_2}
\end{equation}

for the UV divergent terms, thus verifying the Ward Identity (10) for
these terms. Note that the (A5) shows that the four point amplitude is
manifestly O(2) symmetric. So the Ward Identity (10) holds separately for
the UV finite terms (including the Nonplanar terms) as well, although we
are not showing this explicitly. 


\section{One Loop renormalizability of Intermediate shifted theory}

In this section we explicitly compute the  correlation
functions in the shifted theory and  show that the Ward Identity (11) for
this theory is
satisfied, thereby proving the renormalisability of the theory. However we
continue to use $-\mu^2$ as the $(mass)^2$. In the next section we shall
remedy this.
The euclidean form of the noncommutative lagrangian for the shifted theory
is written down below. The star products have been omitted for
convenience. As mentioned earlier we shall treat the shift $v$ as a
background field and finally after taking the limit $\Lambda \rightarrow \infty$, 
restrict $v$ to be a constant field by
putting all momenta on the external $v$ lines to zero. 

\begin{eqnarray}
\L_B &=& \frac{1}{2}(\partial_{\mu}\sigma)^2 + 
\frac{1}{2}(\partial_{\mu}v)^2 + \partial_\mu\sigma\partial^\mu v +
\frac{1}{2}(\partial_\mu\pi)^2  -\mu^2\sigma v
\\  \nonumber &+& [-\frac{1}{2}\mu^2\sigma^2 +
\frac{\lambda}{4}(\sigma \sigma v v +\sigma v \sigma v + \sigma v v \sigma
+vv \sigma \sigma + v \sigma \sigma v + v \sigma v \sigma)]   
\\ \nonumber &+& [-\frac{1}{2}\mu^2\pi^2 + \frac{\lambda}{4}(\pi \pi v v
+ v v \pi \pi)]
+ \frac{\lambda}{4}(v \sigma \sigma \sigma + \sigma \sigma
\sigma v + \sigma \sigma v \sigma + \sigma v \sigma \sigma)
\\ \nonumber &+& \frac{\lambda}{4}\pi \pi \pi \pi + \frac{\lambda}{4}
\sigma\sigma\sigma\sigma
 + [ \frac{\lambda}{4}(\sigma v v v +  v v v
\sigma + v v \sigma v + v \sigma v v )]
 +\frac{\lambda}{4}[\sigma \sigma \pi \pi + \pi \pi \sigma
\sigma ]
\\ \nonumber &+& \frac{\lambda}{4}(\sigma v \pi \pi + v \sigma \pi \pi +
\pi \pi \sigma v + \pi \pi v \sigma] + {\cal L}_{ct}
\end{eqnarray}

In our computation we shall use the symmetric phase results for
the correlation functions with the external $ \phi_1$, $\phi_2 $
lines replaced by $\sigma $, $\pi $, and $v$. With such a way of
computing the correlation  functions of this  theory, the role
played by O(2) symmetry in  the renormalizability of this shifted theory is
transparent. The expressions for the amplitudes corresponding  to those
of the $\sigma$ tadpole and the $\pi$-$\pi$ amplitude are worked out by restricting the background
$v$ field to a constant.

Listed below is the diagramatic representation showing which symmetric
phase process corresponds to which in the shifted theory. The contributions 
to the effective action from each of the symmetric phase processes are
first written down, and then the $\sigma$ and $\pi$-$\pi$ correlation
functions are extracted by finally setting $v$ to a constant field, $v_0$.
The expressions related to a particular diagram are written down
immediately below the diagram. As for the notation, in the expressions for
the effective action, $S_{k}^{m\phi_i n\phi_j}$, the superscript stands
for the process with $m-\phi_i$ and $n-\phi_j$ external lines. The
subscript $k$ stands for the $k$-th contribution to the effective action
$S^{m\phi_i n\phi_j}$ i.e. $S^{m\phi_i n\phi_j}=\Sigma S_{k}^{m\phi_i
n\phi_j}$.

\subsection{$\sigma$ Tadpole Amplitude}


\begin{figure}
\begin{center}
\epsfig{file=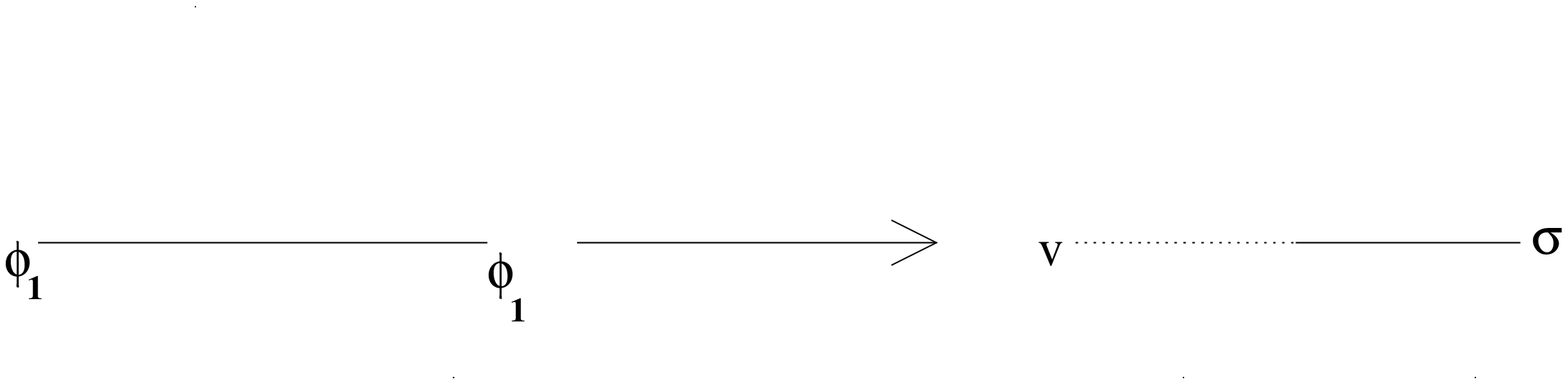, width= 9 cm,angle=0}
\vspace{ .2 in }
\begin{caption}
{   }
\end{caption}
\end{center}
\label{fig1}
\end{figure}

\noindent
\begin{eqnarray}
\ S_1^{2\phi_1}
&=&\int dp_1 dp_2 \frac{1}{2}[ p_1^{\mu}p_{2\mu}+\mu^2]
\phi_1(p_1)\phi_1(p_2)\delta
(p_1+p_2)]
\end{eqnarray}

\begin{equation}
\Gamma_1^{\sigma}= v_0\mu^2
\end{equation}

\begin{figure}
\begin{center}
\epsfig{file=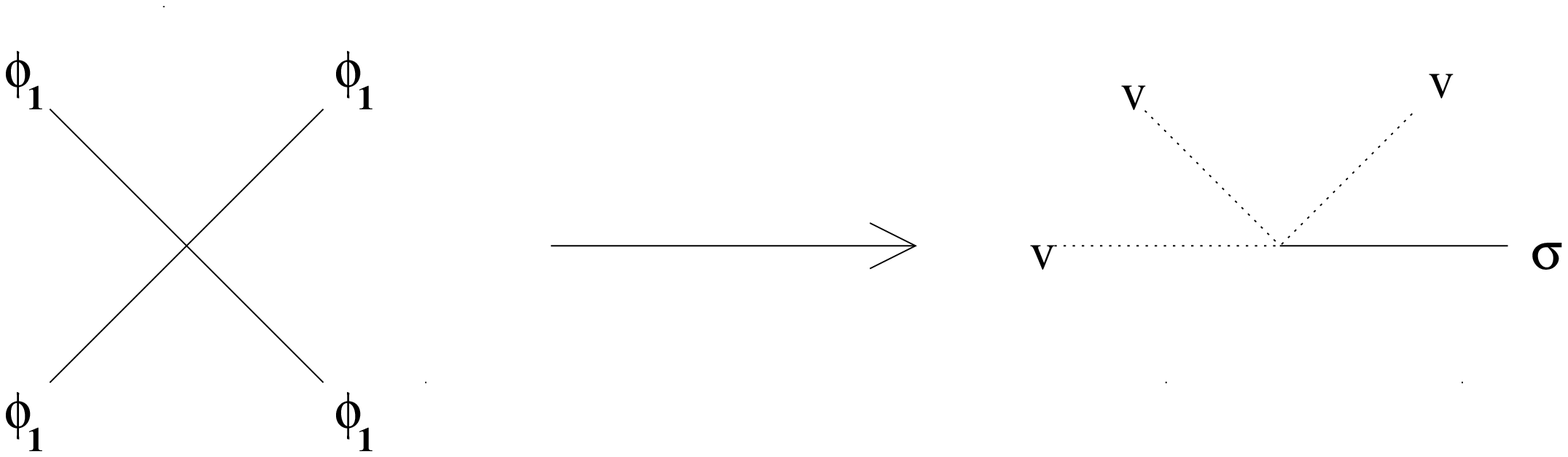, width= 9 cm,angle=0}
\vspace{ .2 in }
\begin{caption}
{   }
\end{caption}
\end{center}
\label{fig1}
\end{figure}

\begin{eqnarray}
\ S_1^{4\phi_1}=\int dp_1 dp_2 dp_3 dp_4
\frac{1}{4!}f_1(p_1,p_2,p_3,p_4)[-2\lambda]\phi_1(p_1)\phi_1(p_2)\phi_1
(p_3)\phi_1(p_4)   
\end{eqnarray}  

\begin{equation}
\Gamma_2^{\sigma}=v_0^3 [-\lambda]
\end{equation}


\begin{figure} 
\begin{center} 
\epsfig{file=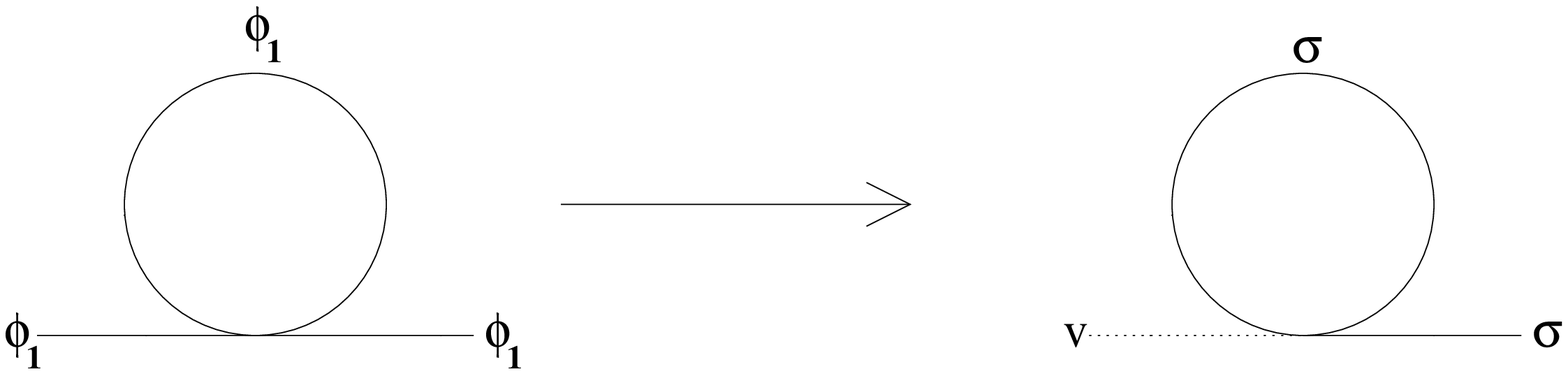, width= 9 cm,angle=0} 
\vspace{ .2 in } 
\begin{caption}
{   }
\end{caption} 
\end{center}  
\label{fig1} 
\end{figure} 

\noindent
\begin{eqnarray}
\ S_2^{2\phi_1} 
&=&\int dp_1 dp_2 \frac{1}{2}[ -2\lambda(\Lambda^2 + \mu^2
ln(\frac{\Lambda^2}{-\mu^2})) + NP+ F] \phi_1(p_1)\phi_1(p_2)\delta
(p_1+p_2)]
\end{eqnarray}

\begin{equation}
\Gamma_3^{\sigma}= v_0[-2\lambda( \Lambda^2 + \mu^2
ln(\frac{\Lambda^2}{-\mu^2})) + IR+ F]
\end{equation}

where NP stands for the nonplanar terms which gives the Infra-Red(IR)
divergent terms  when the external leg $v$ momenta are taken to zero, and
F are the finite pieces.

\begin{figure}
\begin{center}
\epsfig{file=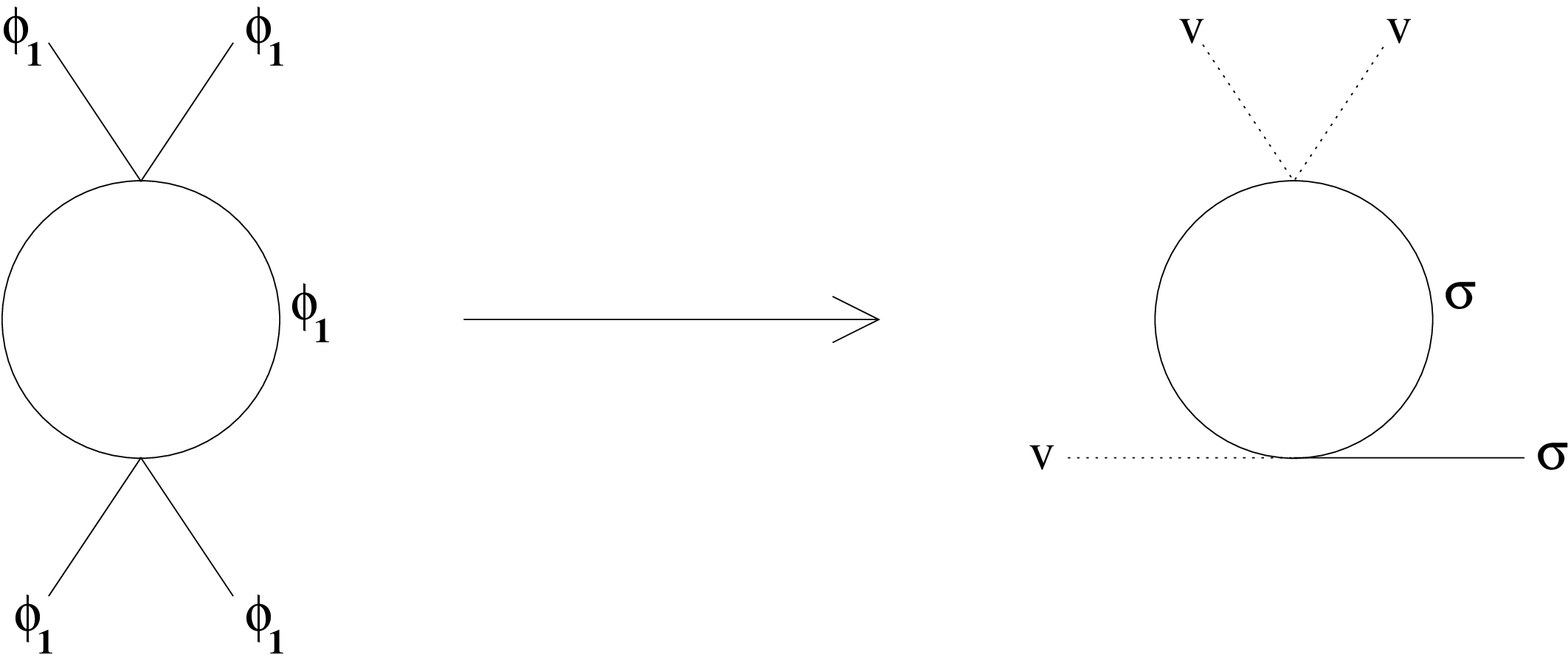, width= 9 cm,angle=0}
\vspace{ .2 in }
\begin{caption}
{   }
\end{caption}
\end{center}
\label{fig1}
\end{figure}

\begin{eqnarray}
\ S_2^{4\phi_1}=\int dp_1 dp_2 dp_3 dp_4
\frac{1}{4!}[f_1(p_1,p_2,p_3,p_4)4\lambda^2ln(\frac{\Lambda^2}{-\mu^2})+NP 
+ F]\phi_1(p_1)\phi_1(p_2)\phi_1
(p_3)\phi_1(p_4)
\end{eqnarray}

\begin{equation}
\Gamma_4^{\sigma}=v_0^3 [2\lambda^2  
ln(\frac{\Lambda^2}{-\mu^2}) + IR+ F]
\end{equation}

\begin{figure}
\begin{center}  
\epsfig{file=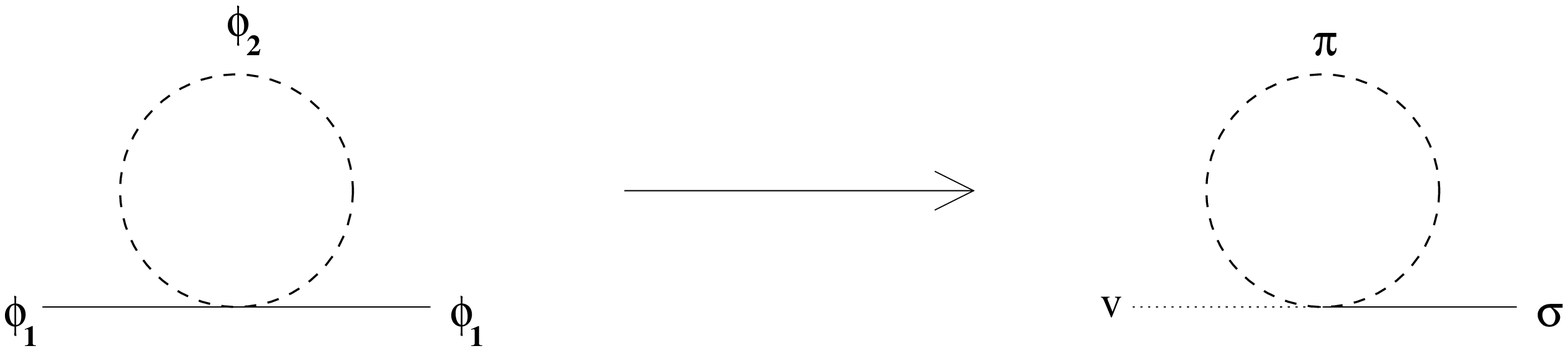, width= 9 cm,angle=0}
\vspace{ .2 in }
\begin{caption}
{   }
\end{caption}
\end{center}
\label{fig1}
\end{figure}

\begin{eqnarray}
\ S_3^{2\phi_1}
&=&\int dp_1 dp_2 \frac{1}{2}[ -\lambda(\Lambda^2 + \mu^2
ln(\frac{\Lambda^2}{-\mu^2})) + NP+ F] \phi_1(p_1)\phi_1(p_2)\delta
(p_1+p_2)]
\end{eqnarray}

\begin{equation}
\Gamma_5^{\sigma}= v_0[-\lambda( \Lambda^2 + \mu^2
ln(\frac{\Lambda^2}{-\mu^2})) + IR+ F]
\end{equation}  

\begin{figure}  
\begin{center}
\epsfig{file=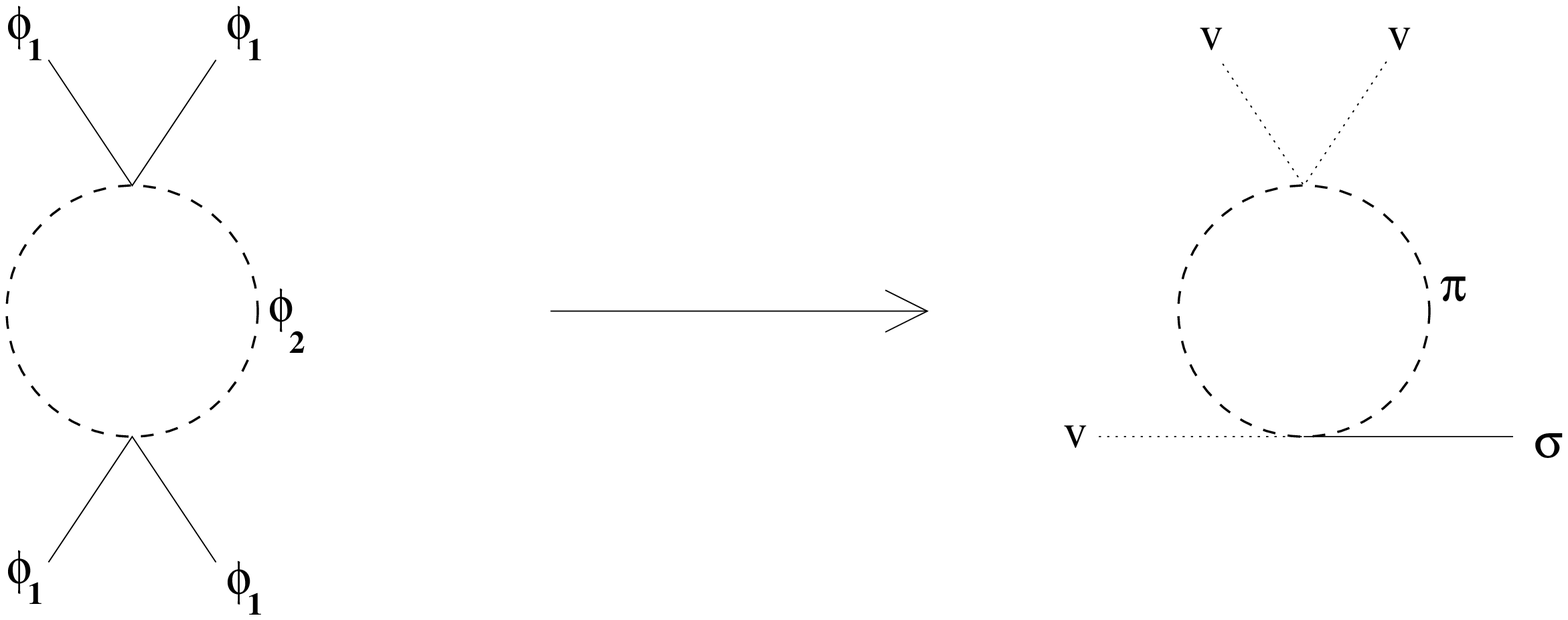, width= 9 cm,angle=0}
\vspace{ .2 in }
\begin{caption}
{   }
\end{caption}
\end{center}
\label{fig1}
\end{figure}

\begin{eqnarray}
\ S_3^{4\phi_1}=\int dp_1 dp_2 dp_3 dp_4
\frac{1}{4!}[f(p_1,p_2,p_3,p_4)\lambda^2ln(\frac{\Lambda^2}{-\mu^2})+NP
+ F]\phi_1(p_1)\phi_1(p_2)\phi_1
(p_3)\phi_1(p_4)
\end{eqnarray}

\begin{equation}
\Gamma_6^{\sigma}=v_0^3 [\frac{1}{2}\lambda^2
ln(\frac{\Lambda^2}{-\mu^2}) + IR+ F]
\end{equation}

Thus the total tadpole amplitude (coming from terms quartic in $\phi^i$ in $\Gamma(\phi^i)$) upto one loop is,
\begin{eqnarray}
\Gamma^\sigma &=&\Sigma \Gamma_i^\sigma \\ \nonumber
&=& v_0\mu^2-\lambda v_{0}^3 + v_0[-3\lambda( \Lambda^2 + \mu^2
ln(\frac{\Lambda^2}{-\mu^2})) + IR+ F]+v_0^3 [\frac{5}{2}\lambda^2
ln(\frac{\Lambda^2}{-\mu^2}) + IR+ F]
\end{eqnarray}

Note that the tree level $\sigma$ tadpole amplitude vanishes when the $v$
field is restricted to a
constant, $v_0$ satisfying $v_{0}^2=\frac{\mu^2}{\lambda}$ as we would have got if we had restricted
$v$ to this constant value at the Lagrangian level itself. This is because
the tree level
amplitudes are totally insensitive to the noncommutativity of the theory. The amplitude is
also Infrared divergent. However this does not affect the UV
renormalisability of the theory.

\subsection{$\pi$-$\pi$ Amplitude}

\begin{figure}
\begin{center}
\epsfig{file=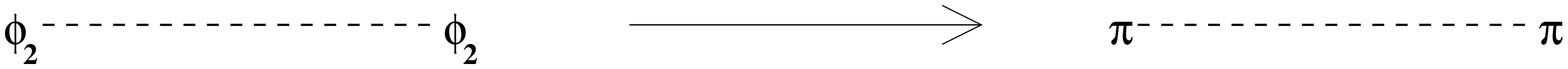, width= 9 cm,angle=0}
\vspace{ .2 in }
\begin{caption}
{   }
\end{caption}
\end{center}
\label{fig1}
\end{figure}

\begin{eqnarray}
\ S_1^{2\phi_2}
&=&\int dp \frac{1}{2}[ -p^2+\mu^2] \phi_2(p)\phi_2(-p)
\end{eqnarray}  

\begin{equation}
\Gamma_1^{\pi\pi}= -p^2+\mu^2
\end{equation}

\begin{figure}
\begin{center}
\epsfig{file=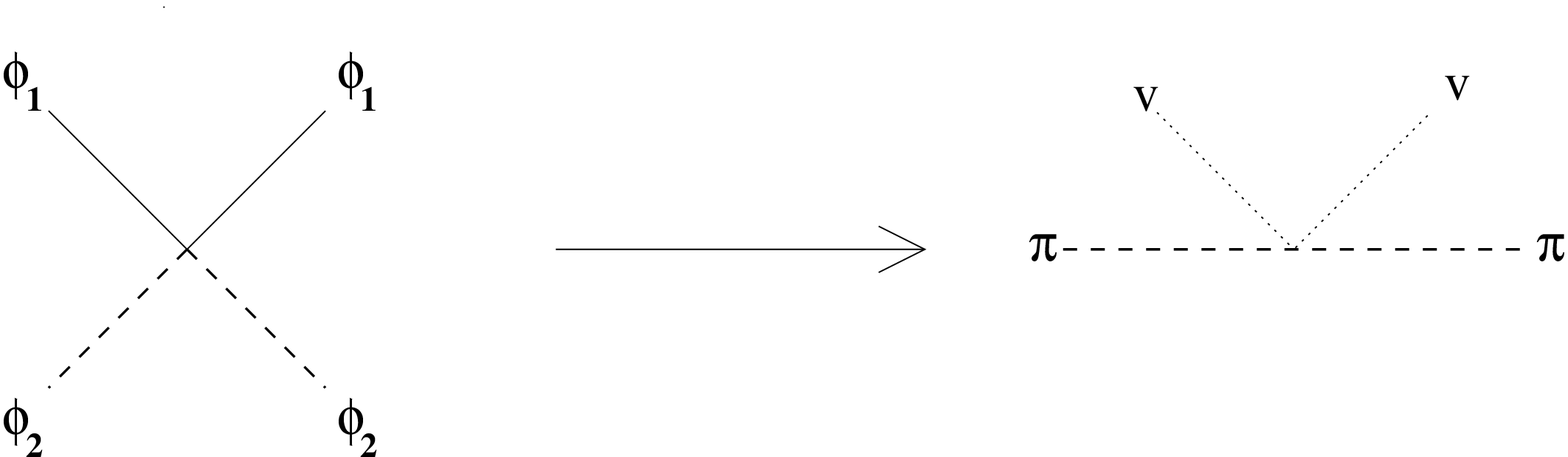, width= 9 cm,angle=0}
\vspace{ .2 in }
\begin{caption}
{   }
\end{caption}
\end{center}
\label{fig1}
\end{figure}

\begin{eqnarray}
\ S_1^{2\phi_1 2\phi_2}=\int dp_1 dp_2 dp_3 dp_4
\frac{1}{4}f_2(p_1,p_2,p_3,p_4)[-2\lambda]\phi_1(p_1)\phi_1(p_2)\phi_2
(p_3)\phi_2(p_4)\delta(p_1+p_2+p_3+p_4)
\end{eqnarray}

\begin{equation}
\Gamma_2^{\pi\pi}=- v_0^2 \lambda
\end{equation}

\begin{figure}
\begin{center}
\epsfig{file=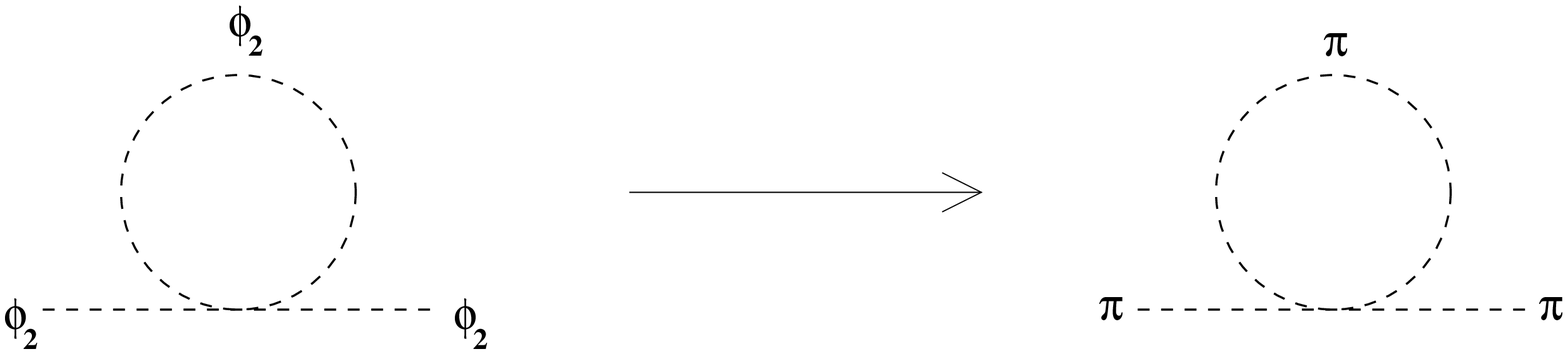, width= 9 cm,angle=0}
\vspace{ .2 in }
\begin{caption}
{   }
\end{caption}
\end{center}  
\label{fig1}
\end{figure}

\begin{eqnarray}
\ S_2^{2\phi_2}
&=&\int dp_1 dp_2 \frac{1}{2}[ -2\lambda(\Lambda^2 + \mu^2
ln(\frac{\Lambda^2}{-\mu^2})) + NP+ F] \phi_2(p_1)\phi_2(p_2)\delta
(p_1+p_2)]
\end{eqnarray}  

\begin{equation}
\Gamma_3^{\pi\pi}= [-2\lambda( \Lambda^2 + \mu^2
ln(\frac{\Lambda^2}{-\mu^2})) + IR+ F]
\end{equation}

\begin{figure}
\begin{center}  
\epsfig{file=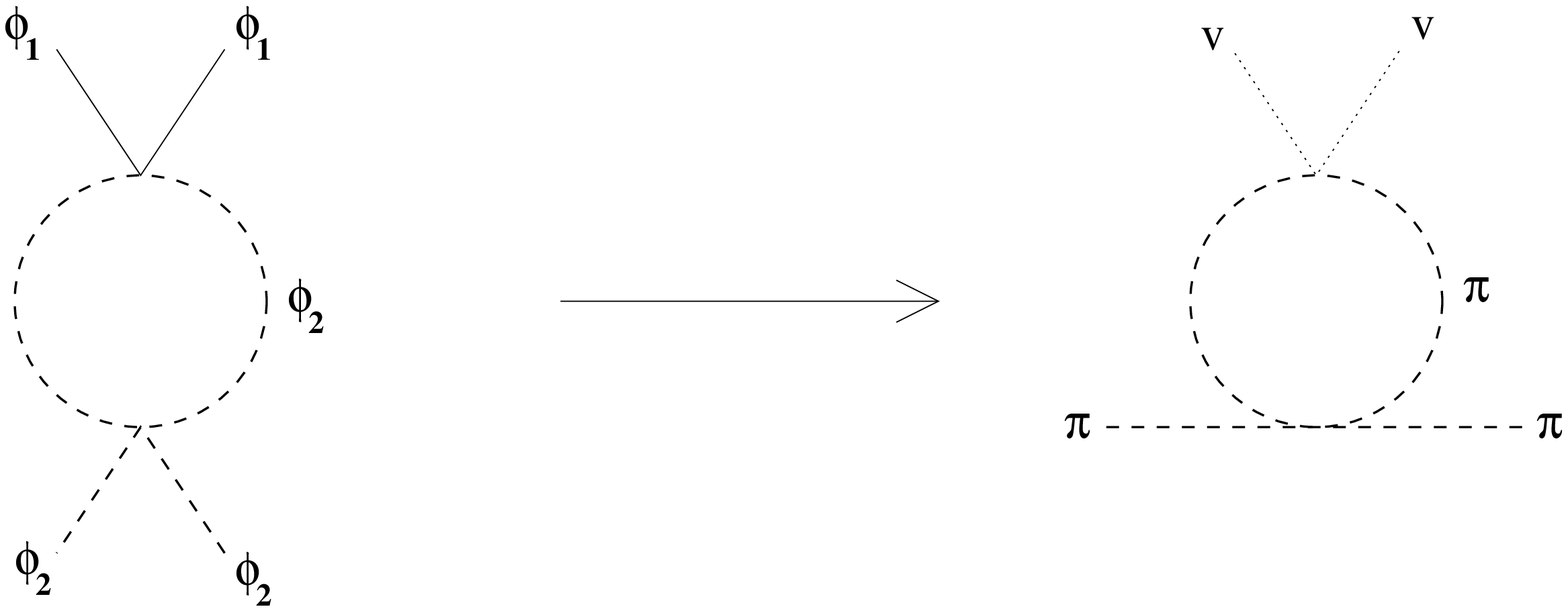, width= 9 cm,angle=0}
\vspace{ .2 in }
\begin{caption}
{   }
\end{caption}
\end{center}
\label{fig1}
\end{figure}

\begin{eqnarray}
\ S_2^{2\phi_1 2\phi_2}=\int dp_1 dp_2 dp_3 dp_4
\frac{1}{4}[f_2(p_1,p_2,p_3,p_4)2\lambda^2ln(\frac{\Lambda^2}{-\mu^2})+NP
+ F]\phi_1(p_1)\phi_1(p_2)\phi_2
(p_3)\phi_2(p_4)
\end{eqnarray}

\begin{equation}
\Gamma_4^{\pi\pi}=v_0^2 [\lambda^2
ln(\frac{\Lambda^2}{-\mu^2}) + IR+ F]
\end{equation}  

\begin{figure}
\begin{center}  
\epsfig{file=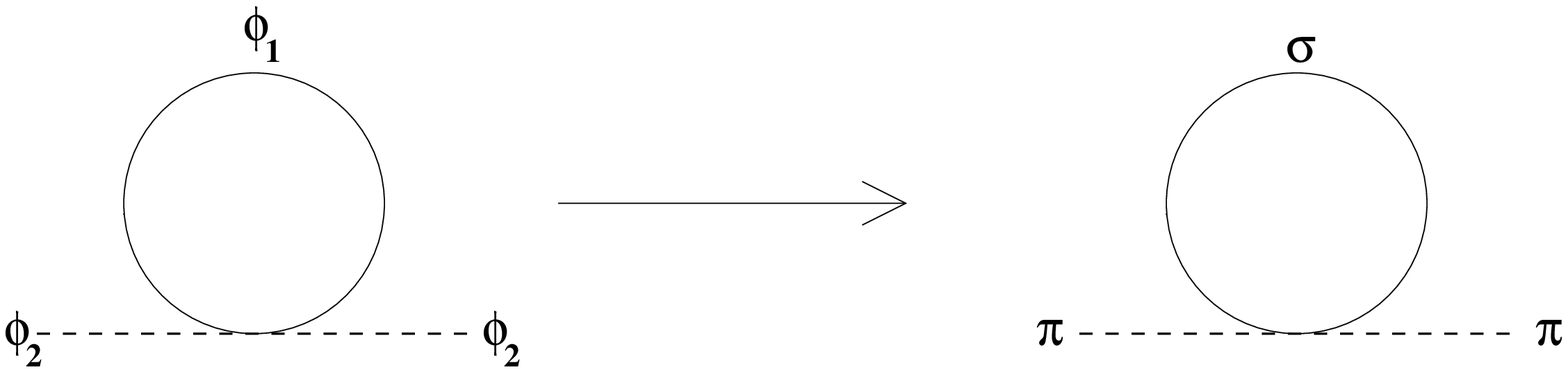, width= 9 cm,angle=0}
\vspace{ .2 in }
\begin{caption}
{   }
\end{caption}  
\end{center}
\label{fig1}
\end{figure}

\begin{eqnarray}
\ S_3^{2\Phi_2}
&=&\int dp_1 dp_2 \frac{1}{2}[ -\lambda(\Lambda^2 + \mu^2
ln(\frac{\Lambda^2}{-\mu^2})) + NP+ F] \phi_2(p_1)\phi_2(p_2)\delta
(p_1+p_2)]
\end{eqnarray}  

\begin{equation}
\Gamma_5^{\pi\pi}= [-\lambda( \Lambda^2 + \mu^2
ln(\frac{\Lambda^2}{-\mu^2})) + IR+ F]
\end{equation}  

\begin{figure}
\begin{center}
\epsfig{file=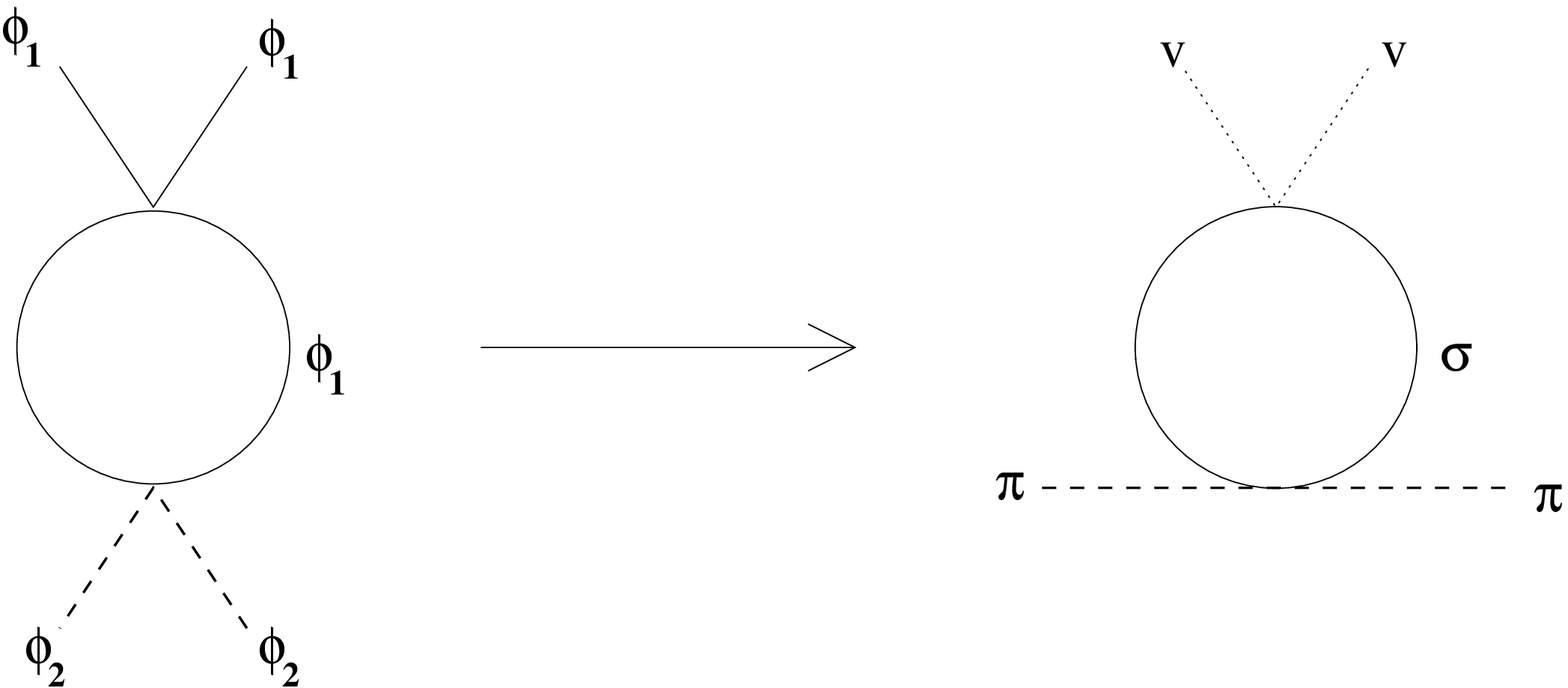, width= 9 cm,angle=0}
\vspace{ .2 in }
\begin{caption}
{   }
\end{caption}
\end{center}
\label{fig1}    
\end{figure}

\begin{eqnarray}
\ S_3^{2\phi_1 2\phi_2}=\int dp_1 dp_2 dp_3 dp_4
\frac{1}{2}[f_2(p_1,p_2,p_3,p_4)\lambda^2ln(\frac{\Lambda^2}{-\mu^2})+NP
+ F]\phi_1(p_1)\phi_1(p_2)\phi_2
(p_3)\phi_2(p_4)
\end{eqnarray}

\begin{equation}
\Gamma_6^{\pi\pi}=v_0^2 [\lambda^2
ln(\frac{\Lambda^2}{-\mu^2}) + IR+ F]
\end{equation}

\begin{figure}
\begin{center}
\epsfig{file=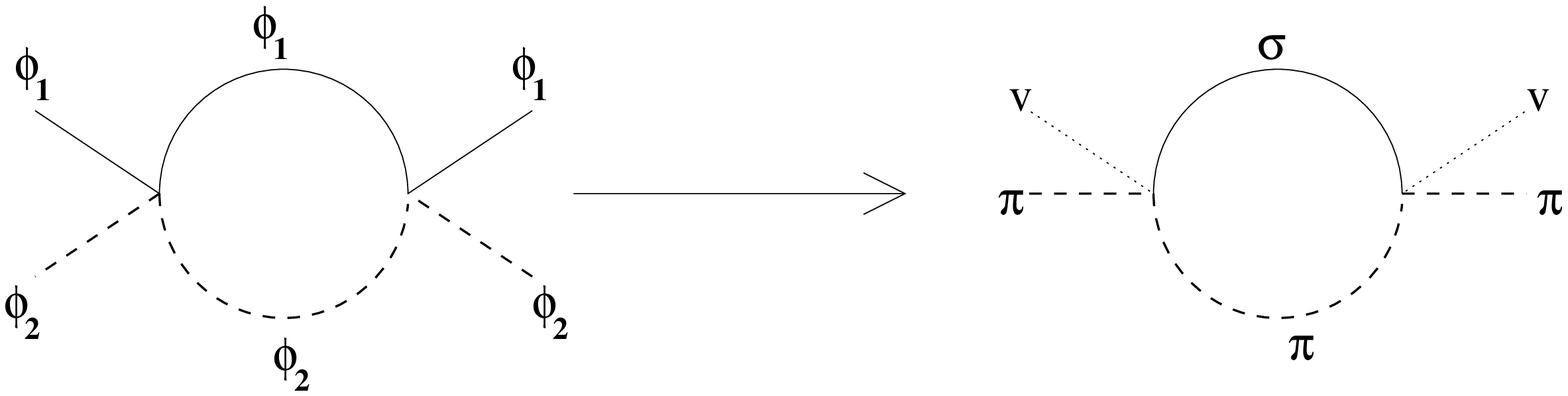, width= 9 cm,angle=0}
\vspace{ .2 in }
\begin{caption}
{   }
\end{caption}
\end{center}
\label{fig1}
\end{figure}

\begin{eqnarray}
\ S_4^{2\phi_1 2\phi_2}=\int dp_1 dp_2 dp_3 dp_4
\frac{1}{4}f_2(p_1,p_2,p_3,p_4)[\lambda^2ln(\frac{\Lambda^2}{-\mu^2})+NP
+ F]\phi_1(p_1)\phi_1(p_2)\phi_2
(p_3)\phi_2(p_4)   
\end{eqnarray}  

\begin{equation}
\Gamma_7^{\pi\pi}=v_0^2 [\frac{1}{2}\lambda^2
ln(\frac{\Lambda^2}{-\mu^2}) + IR+ F]
\end{equation}

Thus upto one loop the total $\pi$-$\pi$ amplitude (coming from terms quartic in $\phi^i$ in $\Gamma(\phi^i)$) is,
 
\begin{eqnarray}
\Gamma^{\pi\pi}&=&\Sigma \Gamma_i^{\pi\pi}\\ \nonumber
&=& -p^2+\mu^2 - \lambda v_{0}^2 + [-3\lambda( \Lambda^2 + \mu^2
ln(\frac{\Lambda^2}{-\mu^2})) + IR+ F] + v_0^2 [\frac{5}{2}\lambda^2
ln(\frac{\Lambda^2}{-\mu^2}) + IR+ F]
\end{eqnarray}

Let us now  analyse these results.
Exactly as in the tadpole the tree level amplitude vanishes when $v$ is
restricted to the
constant value. The amplitude is also IR divergent as we would have expected from the Ward
Identity (11). But even more important are the UV divergent terms. These
are
exactly same as the $\sigma$ tadpole amplitude, equation (34) i.e. 

\begin{equation}
\Gamma^{\sigma} \mid{p=0}=v_0\Gamma^{\pi\pi}\mid{p=0} \nonumber
\end{equation}

for the UV divergent parts. Thus proving the renormalizability of the
theory upto one loop. The way these computations are carried out, the
symmetry of
the unbroken phase makes the renormalizability of this shifted theory
obvious. 
It is apparent from these results that keeping $v$ $x$-dependent 
while taking the continuum limit gives results consistent with the
symmetries of the theory. Since the total theory has O(2) symmetry (action, 
including counterterms and the  measure), the UV finite parts also
should satisfy the WI, although we are not explicitly verifying it.

\noindent
Note that the finite terms for the amplitudes of
this shifted theory (where we have diagrams only with two or four
external legs) are same as that of the shifted commutative
O(2) theory (at one loop level), when the external momenta on the $v$ legs
are set to
zero. This is due to the following reason. A loop integral regulated due
to noncommutativity is \cite{min},

\begin{eqnarray}
\int \frac{d^4 k}{k^2 -\mu^2} e^{ik\wedge p}= \Lambda_{eff}^2 +\mu^2 ln(\frac{\Lambda_{eff}^2}{-\mu^2}) + O(1)
\end{eqnarray}

\noindent
where
\begin{eqnarray}
 \Lambda_{eff}^2=\frac{1}{\frac{1}{\Lambda^2} + pop}
\end{eqnarray}

The nonnommutavity thus only regulates the UV divergent terms and
transforms them into IR divergent terms. The finite terms (O(1)) do not
have any
dependence on the noncommutative parameter, $\theta$ and so can be added
to the finite terms coming from the corresponding Planar graphs. This will
be equal to the finite part from the commutative O(2) theory. 

\noindent
{\it Goldstone Bosons :} The $\pi$-$\pi$ amplitude shows that the UV
divergent terms can be renormalized away. The finite terms vanishes
when the external $\pi$ momentum is taken to zero. As mentioned
these finite terms are the same as that of the commutative theory 
. The $\pi$ field 
is thus massless modulo the IR divergences present in the amplitude 
i.e. the Ward Identity is satisfied formally. The issue is whether
there exists a minimum with $v\neq 0$ or not i.e. is it possible to 
choose $\Gamma^{\sigma}=0$ with $v\neq 0$?  Recall that in 1+1 dimensions there is no 
spontaneous symmetry breaking \cite{coleman} due to untameable IR
divergences in the
broken theory. In our case IR divergences are present in both
phases. Nevertheless an infrared divergent contribution to the $(mass)^2$
may overwhelm the tree level $-\mu^2$ term and result in symmetry
restoration. This can only be decided by more detailed RG analysis.

\section{ Broken theory calculations}

It is clear that working with a tachyonic theory leads to unphysical
values of finite parts arising from the various amplitudes, which are
actually relevant for physical predictions of the theory. One would
naturally like to calculate the amplitudes putting in the expected masses
of the fields in the broken theory. We show in this section how this can
be done without affecting the renormalizability of the Broken Phase
Theory.

Consider the shifted theory lagrangian (21). Now expanding the terms
quadratic
in $v$ such that $v=v_0+v_1$, where $v_0$ is the constant part of $v$, we
have,

\begin{eqnarray}
\L_B &=& \frac{1}{2}(\partial_{\mu}\sigma)^2 +
\frac{1}{2}(\partial_{\mu}v)^2 + \partial_\mu\sigma\partial^\mu v +
\frac{1}{2}(\partial_\mu\pi)^2  -\mu^2\sigma v
\\  \nonumber &+& \frac{1}{2}m_{\sigma}^2 \sigma^2
+ \frac{1}{2}m_{\pi}^2 \pi^2
+ \mbox{ terms containing $v_1$}\\ \nonumber
&+& \frac{\lambda}{4}(v \sigma \sigma \sigma + \sigma \sigma
\sigma v + \sigma \sigma v \sigma + \sigma v \sigma \sigma)
\\ \nonumber &+& \frac{\lambda}{4}\pi \pi \pi \pi + \frac{\lambda}{4}
\sigma\sigma\sigma\sigma
 + [ \frac{\lambda}{4}(\sigma v v v +  v v v
\sigma + v v \sigma v + v \sigma v v )]
 +\frac{\lambda}{4}[\sigma \sigma \pi \pi + \pi \pi \sigma
\sigma ]
\\ \nonumber &+& \frac{\lambda}{4}(\sigma v \pi \pi + v \sigma \pi \pi +
\pi \pi \sigma v + \pi \pi v \sigma] + {\cal L}_{ct}
\end{eqnarray}
 
where,
\begin{eqnarray}
m_{\sigma}^2 &=& 3v_{0}^2\lambda -\mu^2 \\
m_{\pi}^2 &=& v_{0}^2\lambda -\mu^2
\end{eqnarray}

We have seen that the shifted theory is renormalizable for arbitrary value
of $v_0$. Let us assume that $v_0$ is so chosen that the theory is
non-tachyonic and compute the $\sigma$ tadpole and the $\pi$-$\pi$
amplitude. One can do the computations in this broken phase, however we
shall borrow the results from section IV to avoid further
computations. Other than the expansion of terms quadratic in $v$ we shall
follow our previous prescription i.e. setting the $v$ field to a constant 
after all loop computations.

\subsection{$\sigma$ Tadpole Amplitude}
 
The total amplitude to one loop is shown diagramatically below.

\begin{figure}
\begin{center}
\epsfig{file=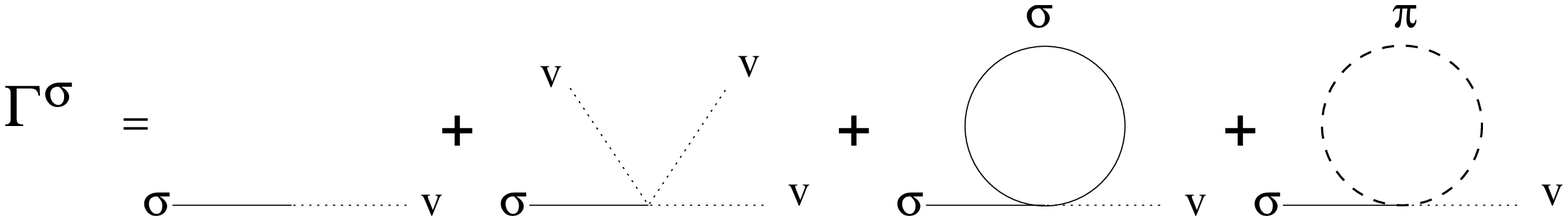, width= 15 cm,angle=0}
\vspace{ .2 in }
\begin{caption}
{   }
\end{caption}
\end{center}
\label{fig1} 
\end{figure}

The terms which would contribute to this amplitude are
from equations (23), (25), (27) and (31) and is shown diagramatically 
in Fig.17. The
other terms of the $\Gamma^{\sigma}$ do not contribute because of
the absence of the quadratic terms in $v$ in this expanded lagrangian (53). The
only difference being the masses. The interaction terms containing $v_1$
which would give rise to diagrams containing external $v_1$ legs would
automatically vanish when we finally restrict $v$ to a constant value of
$v_0$. Taking care of all these facts let us proceed to calculate the  
$\sigma$ tadpole amplitude. 

The component of effective action which would finally contribute to this
amplitude is given by,

\begin{eqnarray}
S_{eff}^{\sigma} &=& \int dp[-p^2+\mu^2]v(p)\sigma(-p)+\int dp_1 dp_2 dp_3
dp_4 
\frac{1}{3!}f_1(p_1,p_2,p_3,p_4)[-2\lambda]\sigma(p_1)v(p_2)v(p_3)v(p_4)
\\ \nonumber &+& \int dp_1 dp_2 [-2\lambda(\Lambda^2
-m_{\sigma}^2ln(\frac{\Lambda^2}{m_{\sigma}^2})) + NP +
F]\sigma(p_1)v(p_2)\delta(p_1+p_2)
\\ \nonumber &+& \int dp_1 dp_2 [-\lambda(\Lambda^2
-m_{\pi}^2ln(\frac{\Lambda^2}{m_{\pi}^2})) + NP +
F]\sigma(p_1)v(p_2)\delta(p_1+p_2)
\end{eqnarray}

\begin{eqnarray}
\Gamma^{\sigma}&=&v_0\mu^2 -v_{0}^3\lambda +v_0[-2\lambda(\Lambda^2
+\mu^2ln(\frac{\Lambda^2}{m_{\sigma}^2})  
-3v_{0}^2\lambda ln(\frac{\Lambda^2}{m_{\sigma}^2}))+IR +F]
\\ \nonumber&+&v_0[-\lambda(\Lambda^2
+\mu^2ln(\frac{\Lambda^2}{m_{\pi}^2})-v_{0}^2 
\lambda ln(\frac{\Lambda^2}{m_{\pi}^2}))+ IR +F]
\end{eqnarray}

Comparing this with equation (34) from section IV we see that
the tree level amplitude and the
quadratic divergence is identical. We have explicitly written down the
masses
$ m_{\sigma}^2(=3v_{0}^2\lambda -\mu^2) $ and $ m_{\pi}^2(=v_{0}^2\lambda
-\mu^2) $ in the coeffecients of the logarithmically divergent terms.
As can be seen in Figures A and B, mass insertions proportional to $v^2$
can change a planar graph into a nonplanar one. Thus when the physical
masses $m_{\sigma}^2$ and $m_{\pi}^2$ are used in the propagators one has
to remember that what looks like a planar graph has some nonplanarity in
it. For the UV divergent terms we prescribe a rule that keeps account of
these. For finite terms it is impossible to do this. Fortunately, for the
finite terms there is no need to keep track of planar and nonplanar
contributions separately.

We divide the $v_{0}^2$ insertions in the $\sigma$
propagators of the UV divergent terms in the ratio of 1:2 corresponding to the division
as we would have got between Planar and Nonplanar terms in the tachyonic
theory thus treating $\frac{1}{3}$ of it as Planar(UV divergent) and
$\frac{2}{3}$ of it as Nonplanar(IR divergent). A similar consideration
with the $\pi$ propagators show that the $v_{0}^2$ insertions there has to be
divided in the ratio of 1:1 as P:NP. 
We explain the reason for this ratio diagramatically below.

\begin{figure}
\begin{center}
\epsfig{file=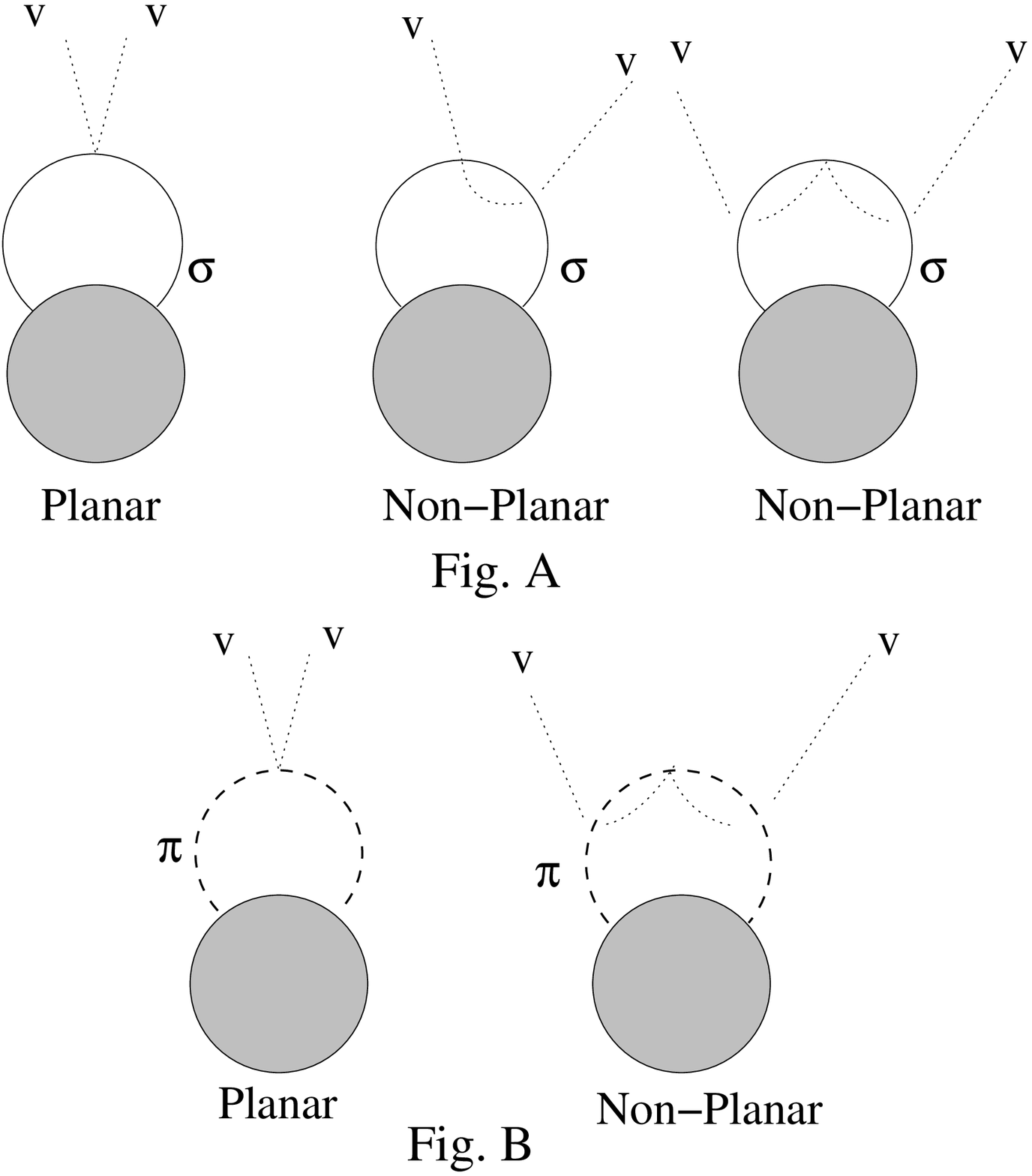, width= 9 cm,angle=0}
\vspace{ .5 in }
\begin{caption}
{ Fig. A gives the rule P:NP = 1:2 for $v^2$ insertions in $\sigma$ propagators and
Fig. B gives the rule P:NP = 1:1 for $\pi$ propagators with $v^2$ insertions  }
\end{caption}
\end{center}  
\label{fig1}
\end{figure}

Each $v^2$ insertion on the propagators would divide the diagram into
Planar and Non-Planar parts as shown in Figures A and B. The ratio in
which these insertions should divide the diagrams in order to preserve
renormalizability of the theory has been worked out by remembering that
the $v$ and the $\sigma$ lines originally correspond to $\phi_1$ lines
in the unbroken theory and the $\pi$ to the $\phi_2$.

With this rule of dividing the UV divergent terms into Planar and Nonplanar
parts, the Planar terms from equation (57) are,

\begin{eqnarray}
\Gamma_P^{\sigma}&=&v_0\mu^2 -v_{0}^3\lambda +v_0[-2\lambda(\Lambda^2
+\mu^2ln(\frac{\Lambda^2}{m_{\sigma}^2})
-v_{0}^2\lambda ln(\frac{\Lambda^2}{m_{\sigma}^2}))]
+v_0[-\lambda(\Lambda^2
+\mu^2ln(\frac{\Lambda^2}{m_{\pi}^2})-\frac{1}{2} v_{0}^2
\lambda ln(\frac{\Lambda^2}{m_{\pi}^2}))]\\ 
&=&v_0\mu^2 -v_{0}^3\lambda +v_0[-3\lambda\Lambda^2
-3\lambda\mu^2ln(\frac{\Lambda^2}{m_{\sigma}^2})
+\frac{5}{2}v_{0}^2\lambda^2 ln(\frac{\Lambda^2}{m_{\sigma}^2}))
+ \frac{1}{2}\lambda^2 v_{0}^2 ln\frac{m_{\sigma}^2}{m_{\pi}^2} 
- \lambda\mu^2 ln\frac{m_{\sigma}^2}{m_{\pi}^2}] 
\end{eqnarray} 

Apart from the tree-level and the UV divergent pieces, equation
(59) contains the $ln(\frac{m_{\sigma}^2}{m_{\pi}^2})$ terms. While
showing
that the Ward Identity (11) holds, we would need to show that these terms
also add up to the same for the $\sigma$ tadpole  and the
$\pi$-$\pi$ amplitudes. We shall not be verifying that the Ward Identity
holds for all the nonplanar terms, but only for the coefficient of
$ln(\frac{m_{\sigma}^2}{m_{\pi}^2})$ terms coming from these nonplanar
terms. Note that the IR terms in equation (57) result from the Nonplanar
terms when we finally set the external momentum on the $v$ legs to
zero.  These Nonplanar terms are of the form,  

\begin{eqnarray}
S_{NP(1)}^{\sigma}=\int dp_1 dp_2[-\lambda(\Lambda_{eff}^2
+\mu^2ln(\frac{\Lambda_{eff}^2}{m_{\sigma}^2})
-3v_{0}^2\lambda ln(\frac{\Lambda_{eff}^2}{m_{\sigma}^2}))]
\sigma(p_1)v(p_2)\delta(p_1+P_2)
\end{eqnarray}

There are other contributions to the Nonplanar terms. These are 
from equation (57), when the $v_{0}^2$ terms are divided into P:NP.
These are,

\begin{eqnarray}
S_{NP(2)}^{\sigma}=\int dp_1 dp_2 [4v_{0}^2\lambda^2
ln(\frac{\Lambda_{eff}^2}{m_{\sigma}^2})
+ \frac{1}{2}v_{0}^2\lambda^2
ln(\frac{\Lambda_{eff}^2}{m_{\pi}^2})]\sigma(p_1)v(p_2)\delta(p_1+P_2)
\end{eqnarray}

\begin{eqnarray}
S_{NP}^{\sigma}&=& S_{NP(1)}^{\sigma} + S_{NP(2)}^{\sigma}
\\ &=&\int dp_1 dp_2 [-\lambda\Lambda_{eff}^2 
-\lambda\mu^2ln(\frac{\Lambda_{eff}^2}{m_{\sigma}^2})
+\frac{15}{2}v_{0}^2\lambda^2 ln(\frac{\Lambda_{eff}^2}{m_{\sigma}^2})
+\frac{1}{2}v_{0}^2\lambda^2 ln(\frac{m_{\sigma}^2}{m_{\pi}^2})]
\sigma(p_1)v(p_2)\delta(p_1+P_2)
\end{eqnarray}

The other finite terms are the same as that of the broken phase of the
commutative O(2) theory as explained earlier. Note that the external
momenta dependences in the $\Lambda_{eff}^2$ in equations (60), (61) may
be
different. However we are only interested in the coefficient of the
$ln(\frac{m_{\sigma}^2}{m_{\pi}^2})$ term which is shown in equation
(63). Now,
taking the contributions from both the planar and the nonplanar parts, the
coefficient of $ln(\frac{m_{\sigma}^2}{m_{\pi}^2})$ in $\sigma$ tadpole
amplitude is, $(\lambda^2 v_{0}^2-\lambda\mu^2)$. 

We now proceed with the computation of the $\pi$-$\pi$ amplitude and show
that this rule can be consistently used to show the renormalizability
of this Broken Phase.

\subsection{$\pi$-$\pi$ Amplitude}

\begin{figure}
\begin{center}
\epsfig{file=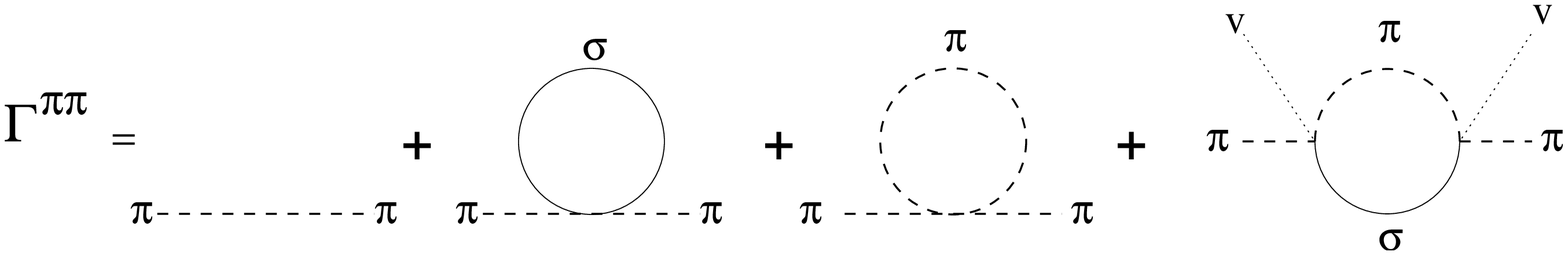, width= 15 cm,angle=0}
\vspace{ .2 in }
\begin{caption}
{   }
\end{caption}
\end{center}
\label{fig1} 
\end{figure}

The terms that will contribute here are from equations (36), (38), (40),
(44) and (48) and is shown diagramatically in Fig.19.

\begin{eqnarray}
S_{eff}^{\pi\pi}&=&\int dp\frac{1}{2}\pi(p)[-p^2+\mu^2-\lambda
v_{0}^2]\pi(-p) 
\\ \nonumber &+& \int dp_1 dp_2\frac{1}{2}[-2\lambda(\Lambda^2-
m_{\pi}^2ln(\frac{\Lambda^2}{m_{\pi}^2})) + NP +
F]\pi(p_1)\pi(p_2)\delta(p_1+p_2)\\ \nonumber
&+& \int dp_1 dp_2\frac{1}{2}[-\lambda(\Lambda^2-
m_{\sigma}^2ln(\frac{\Lambda^2}{m_{\sigma}^2})) + NP +
F]\pi(p_1)\pi(p_2)\delta(p_1+p_2) \\ \nonumber
&+&\int dp_1 dp_2 dp_3 dp_4 \frac{1}{4}
f_2(p_1,p_2,p_3,p_4)[\lambda^2\int_{0}^1 
dx ln(\frac{\Lambda^2}{\Delta^2}) + NP +
F]
\pi(p_1)\pi(p_2)v(p_3)v(p_4)
\end{eqnarray}

\begin{eqnarray}
\Gamma^{\pi\pi}&=&[-p^2+\mu^2-\lambda
v_{0}^2]+[-2\lambda(\Lambda^2+
\mu^2ln(\frac{\Lambda^2}{m_{\pi}^2})  
-\lambda v_{0}^2ln(\frac{\Lambda^2}{m_{\pi}^2})) + IR +F] \\ \nonumber 
&+&[-\lambda(\Lambda^2 +\mu^2ln(\frac{\Lambda^2}{m_{\sigma}^2}) 
-3\lambda v_{0}^2ln(\frac{\Lambda^2}{m_{\sigma}^2}))+IR + F]
\\ \nonumber &+& [\frac{1}{2}\lambda^2
v_{0}^2 \int_{0}^{1} dx ln(\frac{\Lambda^2}{\Delta^2}) +IR +F]
\end{eqnarray}

The expressions under the integral over $x$ is the planar contribution
from the last graph of Fig.19. If $p$ be the total momentum going in and
out of the loop then,

\begin{equation}
-\Delta^2=p^2x(1-x) + x(m_{\sigma}^2-m_{\pi}^2)-m_{\sigma}^2
\end{equation}

After the division of the UV divergent terms into Planar and Nonplanar
terms in appropriate ratios, we have,

\begin{eqnarray}
\Gamma_{P}^{\pi\pi}&=&[-p^2+\mu^2-\lambda
v_{0}^2]+[-2\lambda(\Lambda^2+
\mu^2ln(\frac{\Lambda^2}{m_{\pi}^2})
-\frac{1}{2}\lambda v_{0}^2ln(\frac{\Lambda^2}{m_{\pi}^2}))] \\ \nonumber
&+&[-\lambda(\Lambda^2 +\mu^2ln(\frac{\Lambda^2}{m_{\sigma}^2})
-\lambda v_{0}^2ln(\frac{\Lambda^2}{m_{\sigma}^2}))]
+ [\frac{1}{2}\lambda^2
v_{0}^2ln(\frac{\Lambda^2}{m_{\sigma}^2})] -[\frac{1}{2}\lambda^2
v_{0}^2 \int_{0}^{1} dx ln(\frac{\Delta^2}{m_{\sigma}^2})]\\
\Gamma_{P}^{\pi\pi}\mid_{p=0}&=&[\mu^2-\lambda
v_{0}^2]+[-3\lambda\Lambda^2-
3\lambda\mu^2ln(\frac{\Lambda^2}{m_{\sigma}^2})   
+\frac{5}{2}\lambda^2 v_{0}^2ln(\frac{\Lambda^2}{m_{\sigma}^2})
+ \lambda^2 v_{0}^2 ln(\frac{m_{\sigma}^2}{m_{\pi}^2})
- 2\lambda\mu^2 ln(\frac{m_{\sigma}^2}{m_{\pi}^2})]
-\frac{1}{4}[\lambda^2 v_{0}^2-\lambda\mu^2]
\end{eqnarray}

The last term in the bracket in equation (68) is the contribution from the
integral over $x$, the integration is performed in the limit $p\rightarrow
0$.

We now write down the Nonplanar terms including the contributions
from equation (65) occuring from the division of $v_{0}^2$ terms
in the UV divergent terms. Here again the purpose is to extract the
coefficient of $ln(\frac{m_{\sigma}^2}{m_{\pi}^2})$ term, and although the
various
$\Lambda_{eff}^2$ may have different momenta dependences these terms are
all IR divergent when $v$ is set to a constant and the external $\pi$
momentum is zero.

\begin{eqnarray}
S_{NP(1)}^{\pi\pi}&=&
 \int dp_1 dp_2[-\lambda(\Lambda_{eff}^2
+\mu^2ln(\frac{\Lambda_{eff}^2}{m_{\pi}^2})
-\lambda v_{0}^2ln(\frac{\Lambda_{eff}^2}{m_{\pi}^2}))
+ \frac{7}{2}\lambda^2
v_{0}^2 \int_{0}^{1} 
dx ln(\frac{\Lambda_{eff}^2}{\Delta^2})]\pi(p_1)\pi(p_2)\delta(p_1+p_2)\\
S_{NP(2)}^{\pi\pi}&=&\int dp_1 dp_2 [\lambda^2
v_{0}^2ln(\frac{\Lambda_{eff}^2}{m_{\pi}^2})
+2\lambda^2 v_{0}^2ln(\frac{\Lambda_{eff}^2}{m_{\sigma}^2})]
\pi(p_1)\pi(p_2)\delta(p_1+p_2)
\end{eqnarray}

\begin{eqnarray}
S_{NP}^{\pi\pi}&=&S_{NP(1)}^{\pi\pi}+S_{NP(2)}^{\pi\pi}\\
&=& \int dp_1 dp_2[-\lambda\Lambda_{eff}^2
-\lambda\mu^2ln(\frac{\Lambda_{eff}^2}{m_{\pi}^2})
+\frac{15}{2}\lambda^2 v_{0}^2 ln(\frac{\Lambda_{eff}^2}{m_{\pi}^2})
+ 2\lambda^2 v_{0}^2ln(\frac{m_{\sigma}^2}{m_{\pi}^2})
-\lambda \mu^2
ln(\frac{m_{\sigma}^2}{m_{\pi}^2})\\ \nonumber &-&\frac{7}{2}\lambda^2 
v_{0}^2\int_{0}^{1}
dx ln(\frac{\Delta^2}{m_{\sigma}^2})]\pi(p_1)\pi(p_2)\delta(p_1+p_2) 
\end{eqnarray}

The integral in $x$ can be done with $p\rightarrow 0$, when the
nonplanar terms become IR divergent. The coefficient of the 
$ln(\frac{m_{\sigma}^2}{m_{\pi}^2})$ term from the nonplanar parts is,
$[2\lambda^2 v_{0}^2 -\lambda \mu^2]-[\frac{7}{4}(\lambda^2
v_{0}^2-\lambda\mu^2)]$. Now adding the contributions from both the planar
as well as the nonplanar parts the coefficient is,$(\lambda^2v_{0}^2- 
\lambda\mu^2)$, which is exactly equal to what we had obtained for the
$\sigma$ tadpole amplitude.

Comparing the UV divergences of $\Gamma_{P}^{\sigma}$ equation (59) and
$\Gamma_{P}^{\pi\pi}$ equation (68), the quadratic and the lograthimic 
divergences are identical and thus can be cancelled by the counterterms. This shows
that we have retained the renormalizability of the theory. Note that the
weights of the nonplanar terms in equations (63) and (72) also match
(although
we have not verified equality of these terms with proper momenta
dependences). Furthermore, since the action and the measure are O(2) invariant, as are
the UV divergent pieces, we can conclude that the rest of the UV finite (including IR
divergent) parts of $\Gamma$ satisfy the Ward identity. This would normally imply
the masslessness of the pion. However we do not know whether the IR divergences
imply that the symmetric vacuum has lower energy.

\section{conclusion}

We have shown that for the spontaneously broken noncommutative scalar
field theory with global O(2) symmetry there is no violation of Ward
Identity and the theory is renormalizable to one loop. The violations of
Ward Identity as shown in \cite{kirk} can be traced to the dropping of UV divergent
surface terms. A modification of the prescription, where we treat the
shift $v$ as nonconstant background field and  restrict it to a constant
only after all loop computations, allows us to keep track of all these
surface terms. It is also necessary to keep track of the division of the terms into Planar and
Non-Planar parts. In particular the mass insertions corresponding to the
$v_{0}^2$ terms have to be divided into appropriate proportions to
preserve the renormalizability of the broken theory.

The one loop Infrared divergences seem to provide an infinite positive $(mass)^2$ for all the fields of the theory. 
Thus for finite negative tree level  $(mass)^2$ values, one would conclude that there is symmetry
restoration at one loop. However until one has a way to make sense of the IR divergences this would be quite speculative.
We think this is one of the outstanding issues in this area. Perhaps string theory can provide a clue to
the resolution.  

\newpage
{\bf Acknowledgements :}

\vspace{.2in}
\noindent
We thank N. D. Haridass for useful discussions and careful reading of the manusctipt.

\newpage

\appendix
\section{Feynman Rules for the Symmetric phase}
\noindent
{\bf Propagators :}\\

\begin{figure}
\begin{center}  
\epsfig{file=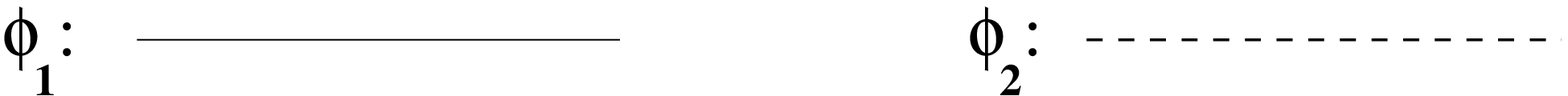, width= 9 cm,angle=0}
\vspace{ .2 in }
\begin{caption}
{   }
\end{caption}
\end{center}
\label{fig1}
\end{figure}

\begin{eqnarray}
\phi_1 : \frac{1}{k^2 -\mu^2}
\\
\phi_2 :\frac{1}{k^2 -\mu^2}
\end{eqnarray}

\noindent
{\bf Interaction Vertex :}\\

\begin{figure}
\begin{center}
\epsfig{file=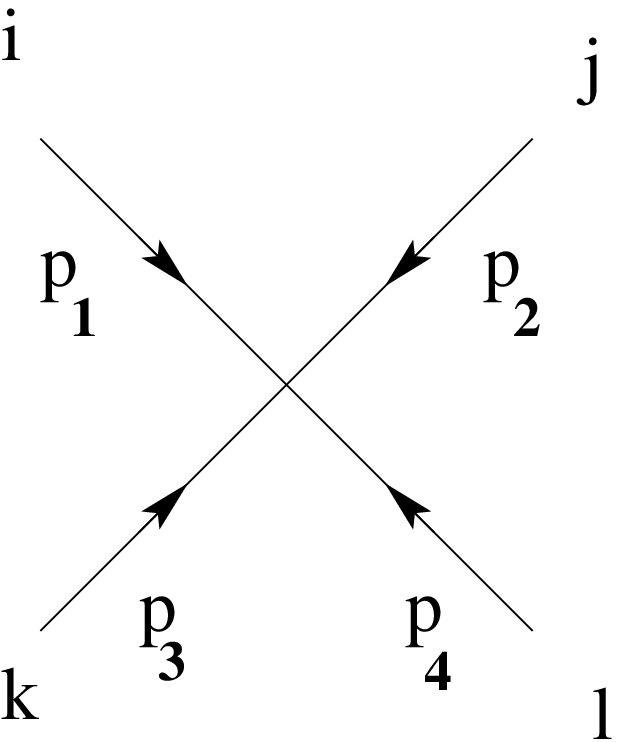, width= 3 cm,angle=0}
\vspace{ .2 in }
\begin{caption}
{   }
\end{caption}
\end{center}
\label{fig1}    
\end{figure}

\begin{eqnarray}
=-2\lambda[\delta^{ij}\delta^{kl}cos(\frac{p_1\wedge
p_2}{2})cos(\frac{p_3\wedge p_4}{2}) +
\delta^{ik}\delta^{jl}cos(\frac{p_1\wedge p_3}{2})cos(\frac{p_2\wedge
p_4}{2}) \\ \nonumber 
&+&  \delta^{il}\delta^{jk}cos(\frac{p_1\wedge p_4}{2})cos(\frac{p_2\wedge
p_3}{2})]
\end{eqnarray}

where,

\begin{equation}
p_i\wedge p_j =p_{i\mu}\theta^{\mu\nu}p_{j\nu}
\end{equation}

\noindent
{\bf Four Point Amplitude :}

\begin{figure}
\begin{center}
\epsfig{file=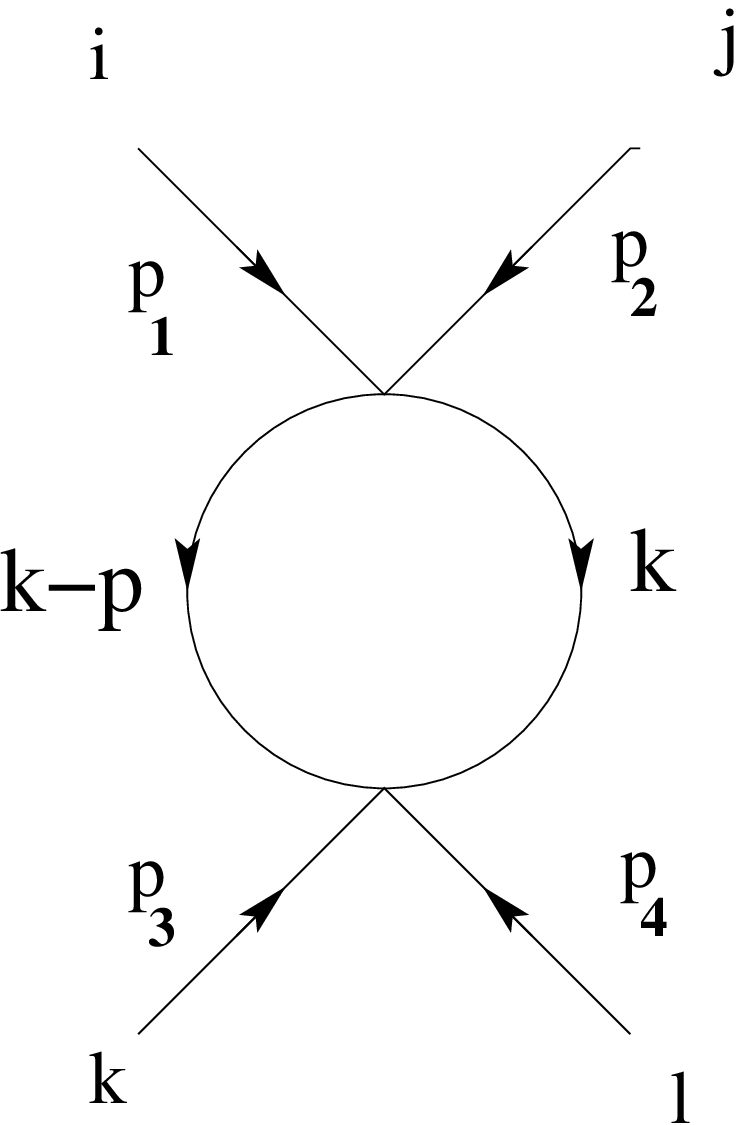, width= 3 cm,angle=0}
\vspace{ .2 in }
\begin{caption}
{   }
\end{caption}
\end{center}
\label{fig1}
\end{figure}

\begin{eqnarray}
=4\lambda^2\frac{1}{2}\int\frac{d^4k}{(2\pi)^4}\frac{V_T(p_1,p_2,p_3,p_4,p,k)} 
{(k^2-\mu^2)[(p-k)^2-\mu^2]}
\end{eqnarray}

Where,

\begin{eqnarray}
V_T(p_1,p_2,p_3,p_4,p,k)&=&[2\delta^{ij}\delta^{kl}[cos^2(\frac{k\wedge
p}{2})cos(\frac{p_1\wedge p_2}{2})cos(\frac{p_3\wedge p_4}{2})\\ \nonumber
&+& cos(\frac{p\wedge k}{2})cos(\frac{p_1\wedge
p_2}{2})cos(\frac{p_3\wedge (p_4+k)}{2})cos(\frac{p_4\wedge k}{2})\\
\nonumber
&+& cos(\frac{p\wedge k}{2})cos(\frac{p_1\wedge
p_2}{2})cos(\frac{p_4\wedge (p_3+k)}{2})cos(\frac{p_3\wedge k}{2})\\
\nonumber
&+& cos(\frac{p\wedge k}{2})cos(\frac{p_3\wedge
p_4}{2})cos(\frac{p_2\wedge (p_1-k)}{2})cos(\frac{p_1\wedge k}{2})\\
\nonumber
&+& cos(\frac{p\wedge k}{2})cos(\frac{p_3\wedge
p_4}{2})cos(\frac{p_1\wedge (p_2-k)}{2})cos(\frac{p_2\wedge k}{2})]\\
\nonumber
&+&\delta^{il}\delta^{jk}[cos(\frac{p_1\wedge k}{2})cos(\frac{p_4\wedge
k}{2})cos(\frac{p_2\wedge (p_1-k)}{2})cos(\frac{p_3\wedge (p_4+k)}{2})\\
\nonumber
&+&cos(\frac{p_2\wedge k}{2})cos(\frac{p_3\wedge
k}{2})cos(\frac{p_1\wedge (p_2-k)}{2})cos(\frac{p_4\wedge (p_3+k)}{2})]\\
\nonumber
&+&\delta^{ik}\delta^{jl}[cos(\frac{p_1\wedge k}{2})cos(\frac{p_3\wedge
k}{2})cos(\frac{p_2\wedge (p_1-k)}{2})cos(\frac{p_4\wedge (p_3+k)}{2})\\
\nonumber
&+&cos(\frac{p_2\wedge k}{2})cos(\frac{p_4\wedge
k}{2})cos(\frac{p_1\wedge (p_2-k)}{2})cos(\frac{p_3\wedge (p_4+k)}{2})]\\
\nonumber
\end{eqnarray}

\begin{equation}
p_1+p_2=p=-(p_3+p_4)
\end{equation}

The other two channels (s and u) can be obtained by interchanging j and k
and j and l and the appropriate momenta.

\end{document}